\documentclass[utf8]{FrontiersinHarvard}

\usepackage{url,hyperref,lineno,microtype,subcaption}
\usepackage{booktabs}
\usepackage{array}
\usepackage{xcolor}
\usepackage{longtable}
\usepackage{amsmath,amssymb}
\usepackage{siunitx}
\usepackage{csquotes}
\usepackage[onehalfspacing]{setspace}

\newcommand{\AnkCcan}{73}
\newcommand{\AnkCroll}{242}

\newcommand{\AnkTauTail}{1876}

\newcommand{\BydArealCath}{2.09}
\newcommand{\BydBoreR}{22.73}
\newcommand{\BydCanDia}{46.3}
\newcommand{\BydCanH}{80.2}
\newcommand{\BydCap}{15.41}
\newcommand{\BydCapErr}{$+$0.51}
\newcommand{\BydCapPred}{15.49}
\newcommand{\BydCathLen}{5.37}

\newcommand{\BydMass}{332}
\newcommand{\BydMassAcc}{340.3}
\newcommand{\BydMassErr}{$-$2.5}
\newcommand{\BydRollR}{22.96}
\newcommand{\BydWindErr}{0.99}
\newcommand{\CanDia}{46}
\newcommand{\CapMeas}{22.078}
\newcommand{\CapMeasSigma}{0.186}
\newcommand{\CapPred}{22.4}
\newcommand{\ChargeErr}{$9.6\times10^{-13}$}

\newcommand{\ElecHeight}{70}
\newcommand{\EnergyErr}{$4.3\times10^{-1}$}

\newcommand{\LamRad}{1.31}
\newcommand{\LfpCcell}{305}

\newcommand{\NParamsE}{21}
\newcommand{\NParamsL}{37}
\newcommand{\NParamsM}{16}
\newcommand{\NParamsTotal}{78}

\newcommand{\NmcMassAcc}{332.8}
\newcommand{\NmcMassErr}{$+$6.5}
\newcommand{\NmcMassMeas}{355.9}
\newcommand{\NmcWindErr}{$+$1.3}

\newcommand{\RCanWallByd}{0.942}
\newcommand{\RCanWallNmc}{0.791}
\newcommand{\RCollectMultiTab}{0.792}

\newcommand{\RCollectTablessM}{0.020}

\newcommand{\RFoilNmc}{0.035}

\newcommand{\ValCrollChange}{$+$12.8}

\newcommand{\ValQMeas}{3.25}

\newcommand{\ValTau}{1812}

\newcommand{\ValTauTrans}{6800}
\newcommand{\ValTauWork}{700}

\newcommand{\WindLen}{3.3}

\newcommand{\BenchTamb}{23.2}
\newcommand{\BenchhA}{0.337}
\newcommand{\BenchH}{22.4}
\newcommand{\BenchC}{277}
\newcommand{\BenchTau}{935}
\newcommand{\BenchRc}{1.34}
\newcommand{\BenchTauc}{4\,740}
\newcommand{\BenchQOne}{15.30}
\newcommand{\BenchQTwo}{14.44}
\newcommand{\BenchQeOne}{14.24}
\newcommand{\BenchQeTwo}{13.41}
\newcommand{\BenchRiseOne}{10.2}
\newcommand{\BenchRiseTwo}{21.9}
\newcommand{\BenchPkOneMeas}{15.0}
\newcommand{\BenchPkTwoMeas}{26.6}
\newcommand{\BenchPkOneMod}{14.1}
\newcommand{\BenchPkTwoMod}{26.6}
\newcommand{\BenchQmOne}{14.10}
\newcommand{\BenchQmTwo}{13.47}
\newcommand{\BenchQfOne}{15.08}
\newcommand{\BenchQfTwo}{14.55}
\newcommand{\BenchTrmsOne}{0.44}
\newcommand{\BenchTrmsTwo}{0.76}
\newcommand{\BenchTrmsTrOne}{0.97}
\newcommand{\BenchTrmsTrTwo}{2.35}
\newcommand{\BenchNxNeg}{20}
\newcommand{\BenchNxSep}{8}
\newcommand{\BenchNxPos}{20}
\newcommand{\BenchConvQ}{less than 0.01}
\newcommand{\BenchConvT}{0.01}
\newcommand{\BenchCoarseQ}{less than 0.01}
\newcommand{\BenchCoarseT}{0.02}
\newcommand{\BenchDsf}{0.77}
\newcommand{\BenchEa}{18.8}
\newcommand{\BenchThetaA}{0.796}
\newcommand{\BenchOcpW}{0.131}
\newcommand{\BenchOcpN}{2.6}

\newcommand{\NmcTauc}{8\,031}
\newcommand{\NmcRc}{0.94}
\newcommand{\NmcEa}{11.0}
\newcommand{\NmcThetaA}{0.700}
\newcommand{\NmcDs}{3.39\times10^{-15}}
\newcommand{\NmcCostTr}{1957}
\newcommand{\NmcCostCal}{9.8}
\newcommand{\NmcNev}{53}

\newcommand{\NmcTrOneTrms}{5.07}
\newcommand{\NmcTrOnePk}{13.5}
\newcommand{\NmcTrOneQ}{19.44}

\newcommand{\NmcTrOnedPk}{$-$7.8}
\newcommand{\NmcTrOnedQ}{$+$0.89}
\newcommand{\NmcTrOnedQabs}{0.89}
\newcommand{\NmcCalOneTrms}{0.52}
\newcommand{\NmcCalOnePk}{21.1}
\newcommand{\NmcCalOneQ}{18.51}

\newcommand{\NmcCalOnedPk}{$-$0.1}
\newcommand{\NmcCalOnedQ}{$-$0.04}

\newcommand{\NmcTrTwoTrms}{8.99}
\newcommand{\NmcTrTwoPk}{29.5}
\newcommand{\NmcTrTwoQ}{17.76}

\newcommand{\NmcTrTwodPk}{$-$14.7}
\newcommand{\NmcTrTwodQ}{$+$1.85}
\newcommand{\NmcTrTwodQabs}{1.85}
\newcommand{\NmcCalTwoTrms}{1.46}
\newcommand{\NmcCalTwoPk}{44.3}
\newcommand{\NmcCalTwoQ}{15.92}

\newcommand{\NmcCalTwodPk}{$+$0.1}
\newcommand{\NmcCalTwodQ}{$+$0.01}

\newcommand{\NmcPkMeasOne}{21.2}
\newcommand{\NmcQMeasOne}{18.55}

\newcommand{\NmcSOCMeasOne}{84}
\newcommand{\NmcPkMeasTwo}{44.2}
\newcommand{\NmcQMeasTwo}{15.91}

\newcommand{\NmcSOCMeasTwo}{72}
\newcommand{\NmcPcanTwo}{1.8}
\newcommand{\NmcPcanOne}{0.46}
\newcommand{\NmcQmeanCalOne}{4.09}
\newcommand{\NmcQmeanCalTwo}{13.2}
\newcommand{\NmcQmeanCalOneTot}{4.5}
\newcommand{\NmcQmeanCalTwoTot}{15.0}

\def\keyFont{\fontsize{8}{11}\helveticabold }
\def\firstAuthorLast{Teodorescu {et~al.}}
\def\Authors{Remus Teodorescu\,$^{1,*}$, Nicolai A. Weinreich\,$^{1}$, Berke Bilgin\,$^{1}$, Yi Zhuang\,$^{1}$ and Wenzel Prochazka\,$^{2}$}
\def\Address{$^{1}$AAU Energy, Aalborg University, Aalborg, Denmark \\
$^{2}$NXP Semiconductors Austria GmbH \& Co KG, Gratkorn, Austria}
\def\corrAuthor{Remus Teodorescu}
\def\corrEmail{ret@et.aau.dk}

\begin{document}
\onecolumn
\firstpage{1}

\title[Parameterisation of NMC and LFP 4680 cells]{Parameterisation of NMC and
LFP 4680 cells for electrochemical and thermal modelling}

\author[\firstAuthorLast ]{\Authors}
\address{}
\correspondance{}
\extraAuth{}

\maketitle

\begin{abstract}
\section{}
An electrochemical--thermal model of the 4680 cylindrical cell is presented: a
Doyle--Fuller--Newman model of the electrode stack coupled to a two-node
thermal balance of the roll and the can (a thermal DFN, TDFN), solved by an
implicit finite-volume scheme. Two cells are parameterised from teardown
geometry and open literature, with every value tagged by provenance: the Tesla
NMC811/graphite tabless cell and the BYD FC4680 LiFePO$_4$ multi-tab cell. The
reaction heat is evaluated at the open-circuit potential of the particle-mean
concentration, which restores the heat of mixing. Both sets are calibrated in the same way: the fixture is
identified from the measurement alone, and a few effective parameters are
adjusted by a Nelder--Mead minimisation of a normalised least-squares cost. For the
NMC cell, against the open 1C and 2C charge test of Ank et al., calibrating
the cathode diffusion time, its activation energy, the anode window and a
contact heat of \NmcRc\,m$\Omega$ outside the voltage path brings the model
within \NmcCalOneTrms\ and \NmcCalTwoTrms\,K of the measured mean can
temperature, the contact heat matching the authors' estimate for their holder. For the LFP cell, discharges at 1C and 2C to 2.0\,V were
measured on the authors' bench; the fixture conductance, \BenchhA\,W\,K$^{-1}$,
follows from the energy balance over a full cycle and the heat capacity,
\BenchC\,J\,K$^{-1}$, from the cooling. The transferred open-circuit potential
of the cathode lacks the end-of-lithiation drop of LiFePO$_4$; with the drop
restored and its width identified from the test, the calibrated set follows
the can temperature within \BenchTrmsOne\,K (1C) and \BenchTrmsTwo\,K (2C) over
the whole discharge and reproduces the peaks, \BenchPkOneMod\ against
\BenchPkOneMeas\,K and \BenchPkTwoMod\ against \BenchPkTwoMeas\,K, and the
relaxation after cut-off. What remains for both cells is what a single can
node and a single representative particle do not describe: the axial gradient
of the NMC can under contact heating, and a gradual decline of the LFP voltage
before the knee at 2C, of up to 90\,mV, which points to the distribution of
particle states in the electrode.

\tiny
\keyFont{\section{Keywords:} lithium-ion battery, 4680 cell, Doyle--Fuller--Newman
model, thermal--electrochemical coupling, parameter identification, LFP, NMC,
bench test}
\end{abstract}

\section*{Nomenclature}

\begin{center}\footnotesize
\setlength{\tabcolsep}{2.4pt}\renewcommand{\arraystretch}{1.12}
\begin{tabular}{@{}lll|lll@{}}\toprule
Symbol & Unit & Description & Symbol & Unit & Description\\\midrule
$a_s$ & m$^{-1}$ & interfacial area per volume & $Q$ & A\,h & charge passed\\
$c_e$ & mol\,m$^{-3}$ & electrolyte concentration & $Q_{s,r}$, $Q_{s,c}$ & W & series-resistance heat in roll, can\\
$\bar c_s$, $c_{se}$ & mol\,m$^{-3}$ & solid: particle mean, surface & $r_s$ & m & particle radius\\
$c_{s,\max}$ & mol\,m$^{-3}$ & maximum solid concentration & $r_i$, $r_o$ & m & roll core and outer radius\\
$c_p$ & J\,kg$^{-1}$\,K$^{-1}$ & specific heat & $R_\mathrm{int}$ & K\,W$^{-1}$ & thermal resistance roll--can\\
$C_r$, $C_c$ & J\,K$^{-1}$ & heat capacity of roll, rest of cell & $R_\mathrm{series}$ & $\Omega$ & series resistance\\
$D_e^\mathrm{eff}$ & m$^2$\,s$^{-1}$ & effective electrolyte diffusivity & $t$ & s & time\\
$D_s$ & m$^2$\,s$^{-1}$ & solid diffusivity & $t_+^0$ & -- & transference number\\
$E_a$ & J\,mol$^{-1}$ & activation energy & $T_r$, $T_c$, $T_\infty$ & K & roll, can, ambient temperature\\
$F$ & C\,mol$^{-1}$ & Faraday constant & $U$ & V & open-circuit potential\\
$h$ & W\,m$^{-2}$\,K$^{-1}$ & can-surface heat-transfer coefficient & $V$ & V & terminal voltage\\
$hA$ & W\,K$^{-1}$ & can--ambient conductance & $w$, $n$ & -- & end-of-lithiation drop: width, exponent\\
$H_e$ & m & electrode height & $x$ & m & coordinate through the stack\\
$i_n$ & A\,m$^{-2}$ & interfacial current density & $\varepsilon_e$, $\varepsilon_s$ & -- & electrolyte, solid volume fraction\\
$I$ & A & applied current & $\theta$, $\theta_0$ & -- & stoichiometry; at 100~\% SOC\\
$j^\mathrm{Li}$ & A\,m$^{-3}$ & volumetric reaction current & $\theta_{\min}$, $\theta_{\max}$ & -- & stoichiometry at 0~\% SOC\\
$J$ & -- & calibration cost & $\kappa^\mathrm{eff}$, $\kappa_D^\mathrm{eff}$ & S\,m$^{-1}$, A\,m$^{-1}$ & electrolyte, diffusional conductivity\\
$L$, $p$ & m & winding length; layer pitch & $\lambda$, $\lambda_r$ & W\,m$^{-1}$\,K$^{-1}$ & thermal conductivity; of the roll\\
$L_\pm$, $L_\mathrm{sep}$ & m & coating, separator thickness & $\rho$ & kg\,m$^{-3}$ & density\\
$m_\mathrm{ref}$ & see Table~\ref{tab:echem} & reaction rate constant & $\sigma^\mathrm{eff}$ & S\,m$^{-1}$ & effective solid conductivity\\
$N$ & mol\,m$^{-2}$\,s$^{-1}$ & molar flux at particle surface & $\tau_c$ & s & cathode diffusion time $r_s^2/D_s$\\
$q$ & W\,m$^{-3}$ & volumetric heat source & $\varphi_s$, $\varphi_e$ & V & solid, electrolyte potential\\
$\bar q$ & mol\,m$^{-4}$ & mean concentration gradient in particle & $\Delta T$, $\Delta Q$, $\Delta V$ & K, A\,h, V & model minus measurement\\
\bottomrule\end{tabular}

\smallskip
\begin{tabular}{@{}ll|ll@{}}\toprule
Abbreviation & Meaning & Abbreviation & Meaning\\\midrule
BPX & Battery Parameter eXchange format & NMC & LiNi$_x$Mn$_y$Co$_z$O$_2$ (NMC811)\\
DFN & Doyle--Fuller--Newman model & OCP & open-circuit potential\\
LFP & LiFePO$_4$ & SOC & state of charge\\
TDFN & thermal DFN & {[M]}, {[L]}, {[D]}, {[E]} & provenance tags (Section~\ref{sec:prov})\\
\bottomrule\end{tabular}
\end{center}

\section{Introduction}
\label{sec:aim}

Physics-based models of lithium-ion cells, from the pseudo-two-dimensional
model of Doyle, Fuller and Newman (DFN) \citep{Doyle1993,Fuller1994} to
multi-scale thermal and electrochemical formulations \citep{Kim2011}, are only
as useful as their parameters. A DFN model of a commercial cell needs several
tens of geometric, transport, kinetic, thermodynamic and thermal parameters,
and most of them cannot be measured on the assembled cell. They are obtained
from teardown and material characterisation
\citep{Ecker2015a,Schmalstieg2018,Chen2020,ORegan2022}, transferred from the
literature \citep{Wang2022LiionDB} or estimated from current and voltage data
\citep{Andersson2022}, and the resulting set is then validated against the
voltage and temperature of the cell \citep{Ecker2015b}. On the thermal side,
lumped models with a core and a surface temperature are established for
cylindrical cells \citep{Forgez2010,Lin2014}.

Two further strands bear on what such a set can be tested against. The heat
generated in a cell follows from an energy balance \citep{Bernardi1985} whose
local form contains a heat of mixing \citep{RaoNewman1997,ThomasNewman2003};
\citet{RichardsonKorotkin2021} derive the exact conservation law for the DFN
model and note that thermal DFN codes commonly omit that term. And terminal
measurements determine only some parameter combinations: the solid-diffusion
parameters enter as $r_s^2/D_s$, and which groups a given set of pulses,
capacities and curves can pin is a question of identifiability
\citep{Forman2012,Andersson2022}. For the 4680 format the measurements are
few. \citet{Ank2023} publish a teardown with open characterisation data, and
\citet{Allgauer2025} report the thermal behaviour of a 4680 cell in cooled
fixtures, including the contact resistance of the cell holder.

Parameter sets are increasingly exchanged between simulation codes in standard
formats such as the Battery Parameter eXchange \citep{BPX2022}. This makes a
set easier to reuse, and also easier to reuse without knowing where its values
came from. \citet{ward2022} describe the general problem for battery data in
the \emph{Principles of the Battery Data Genome}. In their layered metadata
model, simulation model parameters belong to the tertiary tier, and their
quality principle requires that all data and metadata be \enquote{cleaned,
curated, and quality controlled} before they are shared. For measured time
series, existing tools can do this. For a parameter set, in which most values
are not measured directly, it is less clear what quality control means.

A cell-level electrochemical--thermal model needs more than the
electrochemical parameters of a DFN: it needs the geometry of the winding, the
material inventory, the current collection and the thermal network of the
cell, because these set the capacity, the series resistance and the heat path.
No parameter set of this scope has been published for a 4680 cell.
Electrochemical parameters have been published for the NMC811/graphite couple
in the 2170 format \citep{Chen2020}, and teardown studies are now available for
a Tesla 4680 cell with an NMC811/graphite couple \citep{Ank2023} and for the BYD
4680 cell with an LFP/graphite couple \citep{Liu2025}, which allows both cells
to be parameterised on the same basis.
This paper has three aims.

\textbf{First, to identify complete parameter sets for both cells with
traceable provenance.} Every value carries a tag that separates measured data,
literature values transferred from the same chemistry, quantities derived in
this work and our own estimates. The coupled model needs the same
\NParamsTotal\ parameters for both cells, so they are given in one set of
two-column tables with one row per parameter, and rows whose status is
unresolved are marked in red. The Supplementary Material gives the complete
parameterisation of both cells, with the source and the reasoning for each
value.

\textbf{Second, to check each set against quantities that were not used to
build it.} The mass of the assembled cell, its delivered capacity and the
length of the unwound electrode are compared with the values implied by the
parameter set, and the residuals are reported. The mass residual is
\NmcMassErr\,\% for the NMC cell and \BydMassErr\,\% for the LFP cell. The
opposite signs point to hardware that the layer model of the NMC cell does not
contain, rather than to an error common to both identifications
(Section~\ref{sec:audits}).

\textbf{Third, to calibrate the parameter sets against measurement.} The
model is a DFN of the wound cell coupled to its temperature field. Both sets
are taken through the same procedure: the transferred set is compared with a
measurement, the fixture is identified from the measurement alone, and the
effective parameters are calibrated on the voltage and the can temperature at
1C and 2C. For the NMC cell the measurement is the open charge test of
\citet{Ank2023}, with the heat-transfer coefficient of their fixture taken
from the measured cooling (Section~\ref{sec:validation}). For the LFP cell, for which no comparable
measurements are published, the discharge behaviour was predicted before a
bench test and then measured at 1C and 2C, and the set is calibrated on that
test (Section~\ref{sec:bench}).

The DFN describes solid diffusion by Fick's law in a single spherical particle.
This is appropriate for NMC811 but not for the two-phase transformation of
LiFePO$_4$, so the solid diffusivity of the LFP cathode is reported as an
effective parameter (Section~\ref{sec:scopeline}). The bench test shows that the
end of a fast discharge is set by the end-of-lithiation potential of the
cathode, which lies outside the window of the transferred fit, and where the
single-particle description still falls short (Section~\ref{sec:discussion}).

The contributions of this paper are:
\begin{enumerate}
\item a DFN model of a 4680 cell coupled to a two-node thermal balance whose
  heat capacities and internal resistance follow from the parameter set, with
  the fixture boundary taken from measurement;
\item two provenance-tagged parameter sets for real 4680 cells, a tabless NMC
  cell and a multi-tab LFP cell, built from teardown data;
\item the completion of the heat source: a virtual experiment showing that the
  printed source of \citet{GuWang2000} under-predicts the heat when solid
  diffusion is slow, and the evaluation of the reaction heat at the
  open-circuit potential of the particle-mean concentration, consistent with
  \citet{RaoNewman1997}, \citet{ThomasNewman2003} and
  \citet{RichardsonKorotkin2021};
\item a calibration procedure, the same for both cells, that separates
  artefacts of the measurement from the model parameters and ends in a
  Nelder--Mead minimisation of a normalised cost on the voltage and the can
  temperature at 1C and 2C;
\item the calibration of the NMC set on the open charge test of
  \citet{Ank2023}, with a measured boundary condition, which identifies the
  contact heat of the holder and restores the transferred cathode diffusivity
  that a resistance-based heat value had appeared to exclude;
\item a bench test of the LFP cell at 1C and 2C with the fixture conductance
  and heat capacity obtained from the measurement, the calibration of the LFP
  set on it, and the finding that the end of a fast discharge, where the cell
  is hottest, depends on the end-of-lithiation potential of the cathode that a
  transferred fit does not cover;
\item a statement of what remains after calibration, for both cells, and of
  what the open data can and cannot identify.
\end{enumerate}

\section{The two cells}
\label{sec:data}

\subsection{The 4680 format}
\label{sec:format}

The 4680 cylindrical format, \SI{46}{\milli\metre} in diameter and
\SI{80}{\milli\metre} in height, was introduced by Tesla in 2020
\citep{Tesla2020}. Compared with the 2170 cell it succeeds, Tesla stated a
fivefold increase in energy and a sixfold increase in power per cell, a
\SI{16}{\percent} increase in vehicle range and a \SI{14}{\percent} reduction
in cost per kilowatt-hour from the format alone \citep{Tesla2020}. A larger cell means fewer cells per
pack, and therefore fewer cans, welds and connections per kilowatt-hour. The larger diameter was made
possible by tabless current collection. In a conventional cylindrical cell the
current leaves each electrode through one or a few tabs, so the electronic path
and the ohmic heat grow with the length of the electrode. In the tabless design
one uncoated edge of each foil is welded to an end plate along the whole
winding, and the current path is reduced to the width of the electrode
\citep{Tesla2020,Tranter2020}. Tesla presented three cathode families for the
format: iron-based for long cycle life, nickel and manganese based, and high
nickel for long range \citep{Tesla2020}. The first-generation cell
characterised by \citet{Ank2023} uses NMC811.

The \SI{46}{\milli\metre} diameter has since been adopted by other cell and
vehicle manufacturers. BMW has adopted \SI{46}{\milli\metre} cells of
\SI{95}{\milli\metre} and \SI{120}{\milli\metre} height, supplied by CATL and
EVE, for its Neue Klasse vehicles \citep{BMW2022}. LG Energy Solution supplies
4695 cells to Rivian \citep{LGES2024}, and Samsung SDI started production of
46-series cells in 2025 \citep{SamsungSDI2025}. The manufacturers keep the
diameter and vary the height: a taller cell stores more energy without
lengthening the paths for current and heat in the radial direction, and for
tabless cells both the gravimetric and the volumetric energy density increase
with height \citep{Pegel2023}.

Heat removal is the main constraint on the diameter. The surface-to-volume ratio of
a 4680 cell is about half that of a 2170 cell, 0.11 against
\SI{0.22}{\per\milli\metre}, and its electrode is almost five times longer
\citep{Li2023}. For a given layer pitch, the electrode length grows with the
square of the diameter \citep{Baazouzi2023}. Heat generated near the core has
to be conducted either radially through the wound layers, whose through-plane
conductivity is about two orders of magnitude below the axial conductivity of
wound cells \citep{Drake2014}, or axially along the foils to the end faces. Tabless
collection helps in both respects: it reduces the ohmic heat and provides an
axial path from the roll to the end plates, which makes cooling through the
base of the cell effective \citep{Tranter2020,Li2023}. A larger diameter would
again lengthen the radial path from the core to the cooled surfaces
\citep{Tranter2020}, and further increases in energy per cell have instead been
obtained by increasing the height.

The format does not prescribe the chemistry. BYD produces a 4680 cell with an
LFP cathode for both vehicle and stationary storage applications, with safety
and cost as the main motivations \citep{Liu2025}. LFP has a lower specific
energy than NMC811 but contains no nickel or cobalt. The BYD cell
reaches \SI{150.5}{\watt\hour\per\kilo\gram} and
\SI{374.6}{\watt\hour\per\litre} \citep{Liu2025}, against
\SI{232.5}{\watt\hour\per\kilo\gram} and \SI{622.4}{\watt\hour\per\litre} for
the Tesla NMC811 cell \citep{Ank2023}. It also uses a different current
collection, with several tabs on each electrode instead of a tabless design.

\subsection{The two cells studied}
\label{sec:twocells}

The first cell is a Tesla 4680 cell with an NMC811/graphite couple, a liquid
carbonate electrolyte and tabless current collection. The teardown found no
silicon in the anode \citep{Ank2023}. Each foil is left
uncoated along one edge and welded to an end disc along the whole winding
length, so that the current leaves the electrode sideways over tens of
millimetres instead of travelling along its length. The jelly roll is
\CanDia\,mm in diameter, the electrode is approximately \WindLen\,m long and
\ElecHeight\,mm wide, and the capacity measured at C/3 is \CapMeas\,Ah
($\sigma = \CapMeasSigma$\,Ah) \citep{Ank2023}. The DFN model describes this
chemistry without qualification.

The second cell is the BYD FC4680, \BydCanDia\,$\times$\,\BydCanH\,mm and
\BydMass\,g, with an LFP/graphite couple and a nominal voltage of
\SI{3.20}{\volt}, against \SI{3.6}{\volt} for the NMC cell. It stores
\BydCap\,Ah where the NMC cell stores \CapPred\,Ah, about \SI{70}{\percent} of
the charge in the same volume. Its current collection is multi-tab, with four
aluminium tabs on the cathode and three nickel tabs on the anode, one of which
is not connected \citep{Liu2025}. Its geometry, inventory, thermal network and
electrochemistry are identified in the same way as those of the NMC cell, with
the qualification given in Section~\ref{sec:scopeline}.

Both cells are large, so neither the current nor the temperature can be assumed
uniform, and both collect current over a distributed interface rather than at
a point, so the collection topology has to be modelled.
Section~\ref{sec:collection} shows that it changes the collection resistance
by a factor of forty.

\subsection{What transfers between chemistries}
\label{sec:scopeline}

Geometry, material inventory and the thermal and electrical network are
described by the same physics in both cells. A layer stack has a thickness, a
density and a thermal conductivity whatever its active material, the
resistance of a can wall depends only on the steel and the dimensions, and a
collector disk conducts heat to the cap in the same way in both cells. These
quantities are identified and compared directly. The paths of current and heat
differ, however. The model resolves one repeating unit through its thickness
and takes the current as uniform along the winding. This is close to the
tabless cell, whose current leaves each foil over its whole edge, and it is an
approximation for the multi-tab cell, whose current converges on the tabs; the
gradients along the winding are then represented only through the
winding-averaged collection resistance (Section~\ref{sec:collection}) and are
not resolved.

The electrochemistry requires one qualification. LiFePO$_4$ discharges through
a two-phase transformation at nearly constant potential, and the moving phase
boundary inside each particle is not described by the Fickian diffusion in a
single spherical particle that Equation~(\ref{eq:cse}) assumes. The solid
diffusivity of the LFP cathode is therefore reported as an effective parameter
that absorbs the phase-boundary kinetics into a Fickian form, and no physical
meaning is attached to it. With this qualification the DFN needs the same
parameters for both cells, and both sets are given in full.

Three limitations follow for the LFP cell, and each is marked in the tables.
The solid diffusivity is effective, as described above. The model does not
represent the voltage hysteresis between charge and discharge, which is of the
order of tens of millivolts in LFP cells; test~T2 of the bench test measures
it. Finally, \citet{AboutEnergyLFP} report that their parameter set loses
accuracy in voltage and capacity at high rate and low state of charge, which
is the region probed by the rate-retention data of \citet{Liu2025}; the bench
test finds the largest deviation in that region (Section~\ref{sec:bench}). Models
that represent the cathode as an ensemble of phase-separating particles
reproduce the hysteresis \citep{Foster2025,Clarke2026}, and are outside the
scope of this work.

\subsection{Geometry and inventory of both cells}
\label{sec:geoinv}

Table~\ref{tab:echem} gives the geometry of both cells, with each value
carrying the tag it has in its own parameter set, and Supplementary
Table~S5 compares the two cells in full. The geometry of the LFP cell is taken
from the teardown of \citet{Liu2025}, supplemented by quantities derived in
this work where that source gives no value.

Three quantities are not reported by any source: the cathode porosity, the can
wall thickness and the separator thickness. None of them can be obtained from
a single relation. They are determined together from an over-determined system
of mass, capacity and volume balances, described in
Section~\ref{sec:determining}. The outer radius of the roll follows from
Equation~(\ref{eq:spiral}) and agrees with the can bore to within
\BydWindErr\,\% (Section~\ref{sec:audits}).

\subsection{Provenance of the parameters}
\label{sec:prov}

The complete parameter sets of both cells are given in Supplementary
Sections~S7 and~S8: all \NParamsTotal\ values for each cell, each with its unit,
its tag and the reasoning behind it. Every parameter carries one of four tags:

\begin{itemize}
\item \textbf{[M]}: measured or reported for the cell itself, from teardown or
manufacturer data \citep{Ank2023,Liu2025}.
\item \textbf{[L]}: a literature value for the same chemistry, transferred on
the assumption that a 4680 NMC811/graphite electrode behaves like the LG~M50
electrode characterised by \citet{Chen2020}, with the electrolyte correlations
of \citet{Nyman2008}.
\item \textbf{[D]}: derived in this work from tagged quantities by a stated
relation, and therefore carrying their uncertainty.
\item \textbf{[E]}: estimated by the author, with the reasoning recorded.
These values should be treated as fitting parameters.
\end{itemize}

Of the \NParamsTotal\ parameters, \NParamsM\ are \textbf{[M]}, \NParamsL\ are
\textbf{[L]} and \NParamsE\ are \textbf{[E]}. About a third of each set is
therefore estimated, which is to be expected for cells without a published
parameter set, and which is visible only because the tags are printed.

The tags also show where the results are sensitive. The heat-transfer
coefficient at the can surface is \textbf{[E]} and, in the thermal balance of
Section~\ref{sec:twonode}, accounts for \SI{76}{\percent} of the thermal
resistance between the roll and ambient at
$h = \SI{50}{\watt\per\square\metre\per\kelvin}$ and \SI{93}{\percent} in
the test holder of Section~\ref{sec:validation}, so the temperature rise
scales almost linearly with it. The thermal properties of the two coatings and the separator are
also \textbf{[E]}. They set the through-plane conductivity and therefore the
radial temperature gradient, so the shape of the temperature profile depends
on estimates even where its magnitude is constrained by measurement.

The tables are generated from the solver's own objects, and a build-time
assertion fails if any field of any layer lacks a declared source. The tables,
the released metadata and the model therefore remain consistent.

Three choices in the parameter sets require comment.

\textbf{The stoichiometry windows are calibrated.} Transferred with the
open-circuit potentials of \citet{Chen2020}, the stoichiometries at 100\,\%
state of charge let the NMC model discharge well past the capacity of the cell:
to 2.5\,V it delivered \SI{26.1}{\ampere\hour} at C/3, \SI{18}{\percent} above
the \CapMeas\,Ah measured by \citet{Ank2023}. A closure check of this kind is
what the parameter set is built to expose. The windows are therefore calibrated
to the cut-off voltages, a standard step \citep{Birkl2015,Schmalstieg2018}:
with the potentials kept and the electrode capacities fixed by the loadings,
the two 100\,\% stoichiometries are solved so that the open-circuit voltage at
100\,\% is 4.2\,V and a C/20
discharge to 2.5\,V delivers \SI{22.31}{\ampere\hour}, the C/20 capacity of
\citet{Ank2023}. This gives $\theta_0 = 0.7006$ for the anode, against 0.9014
transferred, and 0.2638 for the cathode, against 0.2661; both are tagged
\textbf{[E]}. The anode capacity entering this step is that of the teardown:
the loading of \SI{19.5}{\milli\gram\per\square\centi\metre} at
\SI{1.55}{\gram\per\cubic\centi\metre} gives a coating of
\SI{126}{\micro\metre} per side, tagged \textbf{[M]}, and a solid fraction of
0.70; the set carried an estimated \SI{110}{\micro\metre} before. Over the full
stoichiometry ranges the N/P ratio is then 0.93, an output of the loadings.
Two measurements not used in the calibration test it. The C/3
capacity follows as \SI{22.083}{\ampere\hour}, against \CapMeas\,Ah measured.
The central graphite peak of the differential voltage, which
\citet{Ank2023} place at about \SI{14}{\ampere\hour} of the C/20 charge, falls
at \SI{19.6}{\ampere\hour} in the model. The measured peak corresponds to an
anode stoichiometry of 0.44, whereas the transferred potential places its
central peak at 0.61. That potential is of a graphite--SiO$_x$ electrode, and
\citet{Ank2023} find no silicon in this anode; the shape of the potential, not
the window, is the cause, and it is a finding about the transferred
open-circuit potentials on this cell.
The capacity agreement is therefore evidence only at C/3.

\textbf{Two variants of the maximum solid concentration.} Two pairs of values
are in use for this chemistry, one measured and one obtained by fitting. We use
the measured values for everything that determines capacity, and the fitted
values for the kinetic group, whose members were fitted together and cannot be
separated. This mixed use is marked in the tables.

\textbf{Phase-consistent layer properties.} Density and heat capacity are
averaged over the same inventory, with solid and electrolyte each counted once,
\begin{equation}
\rho_{\mathrm{eff}} = \rho_s\left(1-\varepsilon_e\right) + \rho_e\varepsilon_e,
\qquad
\left(\rho c_p\right)_{\mathrm{eff}}
   = \rho_s c_{p,s}\left(1-\varepsilon_e\right)
   + \rho_e c_{p,e}\varepsilon_e ,
\label{eq:phase}
\end{equation}
where $\varepsilon_e$ is the porosity used in the electrochemical model.
Averaging the two quantities over different inventories, for example a
solid-only density combined with specific heats that include the electrolyte,
changes the thermal mass of the roll by \ValCrollChange\,\%. Such an error does
not affect the voltage, appears directly in the temperature rise, and is not
detected by a mass balance.

\subsection{Closure checks: consistency of a parameter set}
\label{sec:audits}

The quality of a measured time series can be checked by inspection, because a
dropped channel or a clock error leaves a visible trace. A parameter set has
no equivalent. A single value can be plausible, correctly tagged and still
wrong, and the model output it produces gives no independent signal.

A cell that has been torn down has, however, been measured in several
redundant ways that the parameter set must satisfy together. A layer stack
implies a mass; a coating thickness and an areal capacity imply a charge; a
winding pitch and an electrode length imply an outer radius, which must fit
inside a can whose bore was measured separately. None of these quantities is
used to build the set. Each can be computed from the set and compared with the
measurement, and the residual is a measure of quality of the kind that
\citet{ward2022} require for their tertiary tier. We report three such checks
for each cell. For the LFP cell the residuals are those of the initial estimates;
Section~\ref{sec:determining} then uses the same three relations to determine
the quantities that no source reports, after which they close by construction.

\medskip\noindent\textbf{Mass.} Summing all layers over the electrode area, with solid and
electrolyte each counted once through Equation~(\ref{eq:phase}), and adding
the can, the collector disks and the insulators gives the mass of the
assembled cell. For the NMC cell this is \NmcMassAcc\,g against
\NmcMassMeas\,g weighed, a shortfall of \NmcMassErr\,\%. For the LFP cell it is
\BydMassAcc\,g against \BydMass\,g, an excess of \BydMassErr\,\%.

The two residuals have opposite signs. A systematic error in the method, such
as a wrong electrolyte density, a separator counted twice or an error in the
electrode area, would shift both cells in the same direction, so the opposite
signs make such an error unlikely. The shortfall of about 23\,g in the NMC
cell corresponds to hardware that the layer model does not contain: the
mandrel, the tab welds, the vent and burst disc assembly, the crimped cap, and
the electrolyte held outside the roll in the headspace and in the separator
overhang. For the LFP cell, whose teardown reports the mandrel and disk
dimensions, this hardware is largely included. Its excess of about 8\,g comes
from the can wall and separator carried over from the NMC cell
(\SI{0.50}{\milli\metre} and \SI{12}{\micro\metre});
Section~\ref{sec:determining} determines them as \SI{0.418}{\milli\metre} and
\SI{13.7}{\micro\metre}, after which the mass closes.

\medskip\noindent\textbf{Capacity.} For the LFP cell, the reported areal capacity of
\BydArealCath\,mAh\,cm$^{-2}$ over the double-sided cathode gives
\BydCapPred\,Ah against \BydCap\,Ah measured, a residual of \BydCapErr\,\%.
This is an independent prediction. For the NMC cell the transferred windows
delivered \SI{18}{\percent} more than the measured C/3 capacity; the windows
were calibrated to the cut-off voltages and the C/20 capacity
(Section~\ref{sec:prov}), after which the C/3 capacity, not used in the
calibration, closes to \SI{0.02}{\percent}.

\medskip\noindent\textbf{Winding.} The winding does not set the current path of
the tabless cell, which runs across the electrode height, but it sets the outer
radius of the roll and the radial path of the heat, so the check applies to
both cells. An Archimedean spiral of given pitch, core radius and
length has a single outer radius,
\begin{equation}
L = \frac{\pi\left(r_{\mathrm{out}}^{2} - r_{\mathrm{core}}^{2}\right)}{p},
\qquad
r_{\mathrm{out}} = \sqrt{r_{\mathrm{core}}^{2} + \frac{Lp}{\pi}} ,
\label{eq:spiral}
\end{equation}
and this radius must fit inside the can bore. For the LFP cell, the
\BydCathLen\,m cathode at its measured layer pitch gives \BydRollR\,mm against
a bore of \BydBoreR\,mm, an overshoot of \BydWindErr\,\%. For the NMC cell the
same calculation gives \NmcWindErr\,\%.

With the anode coating estimated at \SI{110}{\micro\metre}, as the set carried
before, the NMC winding overshot by \SI{8.8}{\percent}, the largest residual of
the set. Taking the anode thickness from the loading reported in the teardown,
\SI{126}{\micro\metre} per side (Section~\ref{sec:prov}), closes it to
\NmcWindErr\,\%. The check located the error in the one geometric quantity
that had been estimated rather than measured. What remains is within the
variation of layer thickness between the core and the outer wraps that the
micrographs of the teardown show, which a single pitch, as assumed in
Equation~(\ref{eq:spiral}), does not describe.

\medskip\noindent\textbf{Scope of the checks.} The closure checks do not validate the model,
since no dynamics enter them. They test whether the parameter set describes a
physically realisable cell. Their value lies in their ability to discriminate:
checks that were passed equally by both cells would also be consistent with a
residual absorbed into the least constrained parameter. Here the mass check
separates the two cells by sign, and the winding check identified an estimated
anode thickness in the NMC set; with the measured value both cells close the
winding to about one per cent (\NmcWindErr\ and \BydWindErr\,\%).

\subsection{The thermal network of both cells}
\label{sec:networks}

The end hardware of the two cells differs, and the differences partly cancel.
The insulator disk of the LFP cell is \SI{0.29}{\milli\metre} thick, against
\SI{0.65}{\milli\metre} in the NMC cell, which increases the conductance from
the positive collector disk to the can by a factor of 2.2. The multi-tab
construction reduces the conductance from the roll to the collector disk by a
factor of about fifty. The first improves the thermal coupling of the cap, and
the second reduces that of the roll. Both factors are computed from the tagged
geometry of the end hardware and are not adjusted. They are not inputs of the
two-node balance of Section~\ref{sec:twonode}, which lumps the end hardware
into the can node; they describe how a distributed model would divide the heat
between the two ends.

As a result the LFP cell is thermally less asymmetric than the NMC cell. The
decomposition of the LFP network, computed directly rather than scaled from
the NMC result, gives an end-conductance asymmetry ratio of 10.4 against 19.2
and a fixture response coefficient of 2.21 against \SI{2.87}{\kelvin\per\watt}.
Both differences are large enough to matter in an experiment, and neither
could be estimated reliably by hand.

\section{Parameter extraction}
\label{sec:extraction}

The parameters were obtained by three routes: measurement on the cell itself,
transfer from a characterised cell of the same chemistry, and determination
from constraints that several measured quantities impose together. This
section summarises the first two and describes the third, which is used for
the three LFP quantities that no source reports. The group-by-group account,
with the tolerance of every parameter, is given in Supplementary Section~S3
and Supplementary Table~S1.

\subsection{Measured and transferred parameters}
\label{sec:groups}

The tolerance attached to a parameter depends on its tag. Measured
\textbf{[M]} values carry the resolution or scatter of the measurement,
derived \textbf{[D]} values the uncertainty propagated from their inputs, and
estimated \textbf{[E]} values a declared range. Transferred \textbf{[L]}
values carry a transfer band, which is not the precision of the source but
the spread between independent parameterisations of the same material where
two exist.

Four groups determine most of the uncertainty in the predictions. The coating
thicknesses set the capacity linearly,
\begin{equation}
Q = A\,\varepsilon_s L\,c_{s,\max}\,\Delta\theta\,F ,
\label{eq:cap}
\end{equation}
with $A$ the coated area and $\Delta\theta$ the stoichiometry window. Inverting
the areal capacity reported by \citet{Lacey2022} through
Equation~(\ref{eq:cap}) gives a cathode thickness of \SI{82}{\micro\metre}
against the \SI{85}{\micro\metre} measured. The winding length of the NMC cell
has two determinations that differ by \SI{10.7}{\percent}, and it is carried
as a one-sided band. The heat-transfer coefficient at the can surface
accounts for 76 to \SI{93}{\percent} of the thermal resistance between the
roll and ambient, from $h = \SI{50}{\watt\per\square\metre\per\kelvin}$ to
the test holder of Section~\ref{sec:validation} (Equation~\ref{eq:rint}). The electrolyte concentration, in
contrast, has little influence: at \SI{1}{\mole\per\litre} both cells lie close
to the conductivity maximum of the \citet{Nyman2008} correlation.

\subsection{Determination of unmeasured quantities}
\label{sec:determining}

For the LFP cell, \citet{Liu2025} report neither the cathode porosity
$\varepsilon_+$, nor the can wall thickness $w$, nor the separator thickness
$L_{\mathrm{sep}}$. Three relations constrain them. The mass of the assembled
cell is the sum of the can, the end hardware and the layers of the roll,
\begin{equation}
m_{\mathrm{cell}} = m_{\mathrm{can}}(w) + m_{\mathrm{hw}}
  + A\sum_{k} L_k\left[\rho_{s,k}\left(1-\varepsilon_k\right)
  + \rho_e\varepsilon_k\right],
\label{eq:mass}
\end{equation}
where the sum runs over the coatings, the separator and the foils. The
measured areal capacity of the cathode fixes its solid fraction through
Equation~(\ref{eq:cap}). The wound roll must fit the can bore, which for an
Archimedean spiral of length $L_w$ and period $p$ gives
\begin{equation}
\sqrt{r_{\mathrm{core}}^{2} + \frac{L_w\,p}{\pi}}
  = \frac{D_{\mathrm{can}}}{2} - w ,
\qquad
p = L_{\mathrm{Cu}} + L_{\mathrm{Al}} + 2\left(L_- + L_+ + L_{\mathrm{sep}}\right).
\label{eq:bore}
\end{equation}
Equation~(\ref{eq:bore}) contains the wall and the separator but no
chemistry, the capacity relation contains the porosity but no hardware, and
the mass balance contains all three. The system is square and has a unique
solution: $\varepsilon_+ = 0.331$, $w = \SI{0.418}{\milli\metre}$ and
$L_{\mathrm{sep}} = \SI{13.7}{\micro\metre}$. The anode porosity is an input
to the mass balance, not an unknown: inverting the measured anode areal
capacity through Equation~(\ref{eq:cap}) gives 0.319. Before the determination
the cathode porosity had been obtained from the mass balance alone, with a wall
of \SI{0.50}{\milli\metre} taken from the NMC cell, and was 0.419. Re-solving
the system for every input perturbed across its plausible range keeps the wall
between 0.35 and \SI{0.48}{\milli\metre}, the separator between 11.0 and
\SI{16.5}{\micro\metre} and the porosity between 0.29 and 0.37
(Supplementary Table~S2).

Two consequences can be checked independently. The anode porosity of 0.319,
obtained from the anode alone, is close to the 0.331 the three relations give
for the cathode. And the determined wall implies an empty can of
\SI{50}{\gram}, against about \SI{65}{\gram} \textbf{[D]}, computed from the
measured can dimensions for a \SI{0.55}{\milli\metre} wall, the middle of the
0.5 to \SI{0.6}{\milli\metre} case of a structural 4680 can
\citep{BatteryDesign2024}; a single weighing can confirm it.
Because the three relations are used for the determination, they are no longer
independent checks for the LFP cell.

The method applies to quantities that enter a conservation statement: masses,
volumes, charges and lengths. Kinetic, transport and thermal properties enter
none of these relations and have to be measured or transferred.

\subsection{Range of validity of the transferred potentials}
\label{sec:lfpconverge}

The open-circuit potentials of the LFP cell are polynomial and exponential
fits \citep{AboutEnergyLFP}, valid inside the stoichiometry window over which
they were fitted and divergent outside it; the cathode fit returns
\SI{9.06e5}{\volt} at $\theta = 0.05$. A Newton iterate outside the window
therefore produces an unphysical residual. The released code evaluates each
fit inside its window $[\theta_a, \theta_b]$ and extends it linearly outside,
\begin{equation}
U(\theta) = U(\theta_b) + U'(\theta_b)\left(\theta - \theta_b\right),
\qquad \theta > \theta_b ,
\label{eq:guard}
\end{equation}
and correspondingly below $\theta_a$, which leaves the potential unchanged
inside the window and keeps a gradient outside it. The $\tanh$ forms of
\citet{Chen2020} used for the NMC cell are bounded and need no such extension.

\section{The model}
\label{sec:fdm}

\begin{figure}[h!]
\begin{center}
\includegraphics[width=0.98\textwidth]{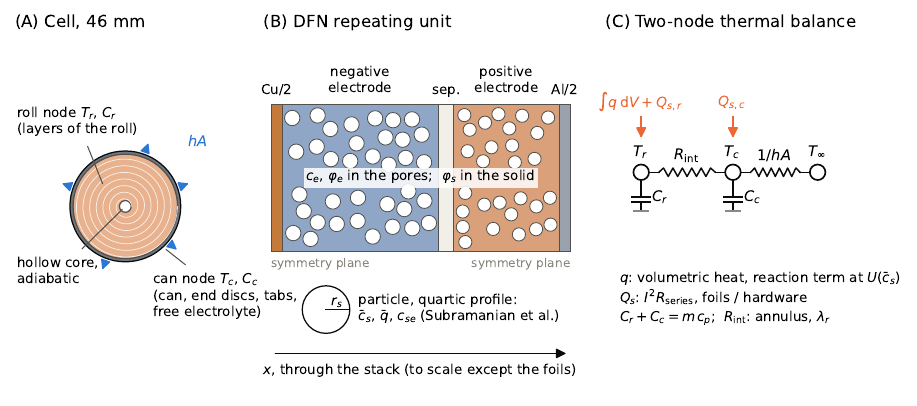}
\end{center}
\caption{The model. \textbf{(A)}~Cross-section of the 4680 cell: the jelly
roll around its hollow core and the can. The roll node carries the heat
capacity of the layers of the roll at their mean temperature $T_r$; the can
node carries the rest of the cell, the can, end discs, tabs and electrolyte
outside the roll, at $T_c$, and exchanges heat with the surroundings through
$hA$ over the whole exterior. \textbf{(B)}~The repeating unit of the DFN
model, resolved through its thickness from the centreline of the copper foil
to that of the aluminium foil; both boundaries are symmetry planes because the
electrodes are coated on both faces. The particle concentration follows the
quartic profile of \citet{Subramanian2005}. \textbf{(C)}~The two-node thermal
balance of Equation~(\ref{eq:energy}), with the heat of the electrode stack and
the $I^2R_{\mathrm{series}}$ heat entering the two nodes and the internal
resistance of Equation~(\ref{eq:rint}).}
\label{fig:model}
\end{figure}

\begin{figure}[h!]
\begin{center}
\includegraphics[width=0.98\textwidth]{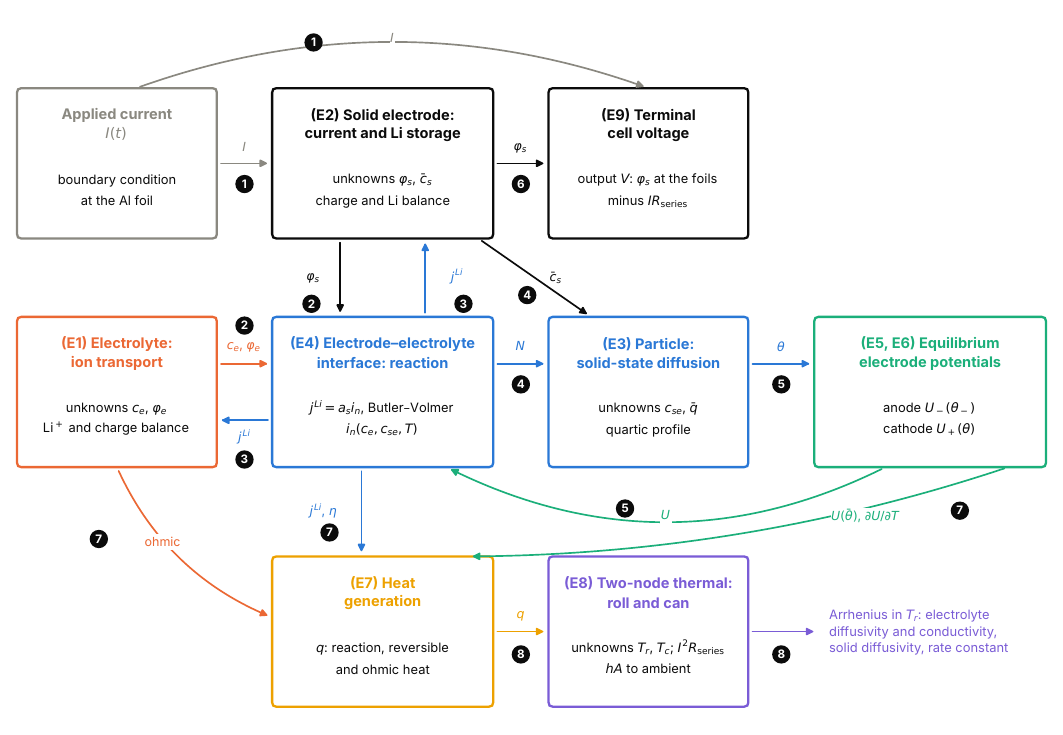}
\end{center}
\caption{Coupling of the equations of the TDFN, labelled as in
Table~\ref{tab:equations}. Boxes are the equations and their unknowns, arrows
the quantities passed between them, and the numbers the steps described in
Appendix~A.1.}
\label{fig:couple}
\end{figure}

The model is the porous-electrode formulation of \citet{Doyle1993} and
\citet{Fuller1994}, extended to non-isothermal operation by
\citet{GuWang2000}, applied to one repeating unit of the wound electrode and
coupled to a two-node thermal balance of the cell (Figure~\ref{fig:model}).
We refer to it as a thermal DFN (TDFN). The transport and charge balances and
the Butler--Volmer kinetics are those of the textbook DFN; the complete set of
equations, labelled (E1) to (E9), is given in Appendix~A
(Table~\ref{tab:equations}), with their coupling (Figure~\ref{fig:couple}).
This section describes only what differs from the textbook form: the domain
of the wound cell, the particle closure, the heat source, the two-node thermal
balance and the current collection.

\subsection{Domain and particle closure}
\label{sec:domain}

The repeating unit is resolved through its thickness $x$, from the centreline
of the copper foil to that of the aluminium foil; the electrode is coated on
both faces, so the boundaries are symmetry planes and the unit stands for the
whole winding with the current taken as uniform along it. Each control volume of an electrode carries one representative particle of
radius $r_s$ and mean concentration $\bar{c}_s$. Instead of resolving the
diffusion inside it, its surface concentration $c_{se}$ follows the quartic profile of
\citet{Subramanian2005}, which adds the volume-averaged concentration gradient
$\bar{q}$ as one state per control volume,
\begin{equation}
\frac{\mathrm{d}\bar{q}}{\mathrm{d}t}
  = -\frac{30 D_s}{r_s^2}\,\bar{q} - \frac{45}{2}\,\frac{N}{r_s^2},
\qquad
\frac{35 D_s}{r_s}\left(c_{se}-\bar{c}_s\right) - 8 D_s \bar{q} = -N,
\qquad a_s = \frac{3\varepsilon_s}{r_s},
\label{eq:cse}
\end{equation}
where $N = \pm j^{\mathrm{Li}}/(a_s F)$ is the molar flux out of the particle
surface, so that the surface responds to a step in current partly at once and partly
with the time constant $r_s^2/(30 D_s)$, and reduces at constant current to the
parabolic closure $c_{se}-\bar{c}_s = -N r_s/(5D_s)$ of \citet{GuWang2000}.

\subsection{Heat source}
\label{sec:heat}

With the reaction current $j^{\mathrm{Li}}=a_si_n$ of Butler--Volmer kinetics
(Appendix~A), heat is generated at the volumetric rate
\begin{equation}
q = a_s i_n\!\left(\varphi_s-\varphi_e-U+T\frac{\partial U}{\partial T}\right)
  + \sigma^{\mathrm{eff}}\left(\frac{\partial\varphi_s}{\partial x}\right)^{2}
  + \kappa^{\mathrm{eff}}\left(\frac{\partial\varphi_e}{\partial x}\right)^{2}
  + \kappa_D^{\mathrm{eff}}\,\frac{\partial\ln c_e}{\partial x}\,
    \frac{\partial\varphi_e}{\partial x} .
\label{eq:q}
\end{equation}
Equation~(\ref{eq:q}) is the volumetric heat source of \citet{GuWang2000},
with one completion. As printed, it evaluates $U$ at the particle surface and
neglects the heat of mixing released as concentration gradients inside the
particles relax \citep{RaoNewman1997,ThomasNewman2003,RichardsonKorotkin2021}.
The model evaluates the reaction heat at the open-circuit potential of the
particle-mean concentration, which restores this term; Butler--Volmer
kinetics keep $U$ at the surface. This is the distributed counterpart of the
cell-level irreversible heat $I(U_{\mathrm{avg}} - V)$ of \citet{Bernardi1985},
not a new heat model. It is exact for the heat integrated until the particles
have relaxed; while a gradient is building up, as after a step in current, it
counts as heat the energy still stored in that gradient, which the mixing term
releases only later. The entropic term uses the coefficients of
\citet{ORegan2022}. For the NMC cell at 1C and 50\,\% state of charge the
completed source agrees with a resolved-particle model including the heat of
mixing \citep{Sulzer2021} to within \SI{0.26}{\watt} for diffusion times
between 680 and \SI{6800}{\second}. With the surface form, the heat is nearly
independent of the solid diffusivity (3.82 to \SI{3.95}{\watt} over the same
range, against 4.13 to \SI{5.86}{\watt} completed), so it would appear to agree
with a measured heat by omission (Supplementary Table~S6).
The electrolyte diffusivity and conductivity, the solid diffusivity and the
rate constant follow Arrhenius laws \citep{Doyle1996} in the temperature of the
roll. The open-circuit potentials and the electrolyte correlations are given
in Supplementary Section~S2.

\subsection{Thermal balance}
\label{sec:twonode}

The temperature of the cell is carried by two nodes \citep{Forgez2010,Lin2014}:
the roll, at its heat-capacity-weighted mean temperature $T_r$, and the can,
at $T_c$,
\begin{equation}
C_r\frac{\mathrm{d}T_r}{\mathrm{d}t}
  = \int_{\mathrm{roll}} q\,\mathrm{d}V + Q_{s,r} - \frac{T_r-T_c}{R_{\mathrm{int}}},
\qquad
C_c\frac{\mathrm{d}T_c}{\mathrm{d}t}
  = \frac{T_r-T_c}{R_{\mathrm{int}}} + Q_{s,c} - hA\left(T_c-T_\infty\right).
\label{eq:energy}
\end{equation}
$C_r$ is the heat capacity of the layers of the roll and $C_c$ the rest of the
cell, the can, the end discs, the tabs and the electrolyte outside the roll,
so that $C_r + C_c = m c_p$. The series resistance of
Section~\ref{sec:collection} is part of the cell's loss: its heat
$I^2R_{\mathrm{series}}$ enters as $Q_{s,r}$ for the share of the foils and as
$Q_{s,c}$ for the tabs, welds and end discs. The internal resistance is the
exact mean-to-surface resistance of a uniformly heated annulus of height $H_e$
between the adiabatic core $r_i$ and the roll surface $r_o$,
\begin{equation}
R_{\mathrm{int}} = \frac{1}{2\pi\lambda_r H_e}\,
  \frac{\tfrac14\left(r_o^2-r_i^2\right)^{2}
        - 2r_i^2\left[\tfrac14\left(r_o^2-r_i^2\right)
        - \tfrac12 r_i^2\ln\left(r_o/r_i\right)\right]}
       {\left(r_o^2-r_i^2\right)^{2}},
\label{eq:rint}
\end{equation}
which reduces to $1/(8\pi\lambda_r H_e)$ for a solid cylinder; $\lambda_r$ is
the through-stack conductivity of the roll, the layers in series
(\LamRad\,W\,m$^{-1}$\,K$^{-1}$ for the NMC cell). The can exchanges heat with
its surroundings through $hA$, over the whole exterior. The electrochemistry
sees $T_r$; a sensor on the can reads $T_c$.

\subsection{Current collection}
\label{sec:collection}

Current collection enters the model through the series resistance. The NMC cell
collects tablessly, over the whole edge of each foil into an end disc; the LFP
cell through four aluminium tabs on the cathode, at radii of 0.4, 13.1, 16.0
and \SI{18.3}{\milli\metre}, and three nickel tabs on the anode, at 9.0, 15.6
and \SI{20.0}{\milli\metre}, one of which carries no current \citep{Liu2025}.
Averaged over the winding, the collection resistance is
\RCollectTablessM\,m$\Omega$ for the tabless cell and \RCollectMultiTab\,m$\Omega$
for the multi-tab cell.\footnote{This is the resistance of the path from a
wrap, through the end disc or the tabs, to the terminal, averaged over the
winding. It is not the foil resistance of \RFoilNmc\,m$\Omega$ listed for the
NMC cell in Supplementary Table~S4, which is the path across
the electrode height inside the foils. In the model the NMC series resistance
is the foil resistance plus the estimated hardware resistance, which contains
the end-disc path; the LFP series resistance is the winding-averaged collection
resistance plus the can wall and the hardware.} The can walls of the two cells have nearly equal
resistances, \RCanWallNmc\ and \RCanWallByd\,m$\Omega$. In the tabless cell the
collection resistance is two orders of magnitude below the can wall and can be
neglected in a lumped description; in the multi-tab cell the two are of the
same size.

\subsection{Discretisation and verification}
\label{sec:discretisation}

The repeating unit is discretised by the cell-centred finite-volume method on
a grid that conforms to the layers, so that no control volume contains two
materials. Face coefficients are harmonic means, which is required because the
foils are seven orders of magnitude more conductive than the adjacent
electrodes, and each foil is resolved by one control volume. Time is advanced
by backward Euler with one Newton solve over all unknowns per step, and each
linear system is equilibrated before a sparse LU factorisation; the can node
of Equation~(\ref{eq:energy}) is eliminated exactly inside each step.
Supplementary Section~S1 gives the details. The grid used
for all results has one control volume in each foil, \BenchNxNeg\ in the
negative electrode, \BenchNxSep\ in the separator and \BenchNxPos\ in the
positive electrode (Figure~\ref{fig:model}B). Doubling the number of control
volumes in each layer and halving the voltage step that controls the time step
changes the delivered charge of the LFP model of the diagnostic calibration
(Section~\ref{sec:bench}) by \BenchConvQ\,\% and
the peak can temperature by \BenchConvT\,K at 2C; halving them changes the same
quantities by \BenchCoarseQ\,\% and \BenchCoarseT\,K. The identified
parameters therefore do not depend on the grid at this resolution.

The integral of the interfacial current over each electrode matches the
applied current to \ChargeErr\ in relative terms, and the integral of
Equation~(\ref{eq:q}) matches the electrical dissipation to \EnergyErr. Over a
2C charge in the test holder of Section~\ref{sec:validation}, the heat
generated equals the heat stored in the two nodes plus the heat lost to the
surroundings to within \SI{0.3}{\percent}. The solver reproduces the delivered
capacity reported by \citet{GuWang2000} for their cell to within two
percentage points of depth of discharge in the coupled, decoupled and
isothermal cases.

\section{Comparison with measurement}
\label{sec:validation}

Both parameter sets are compared with measurement in the same way, in the
five steps of Figure~\ref{fig:triangulation}: (1)~the set is transferred from
published values, each tagged, and its prediction is stated; (2)~the
measurement is cross-checked against other published data on quantities the
model does not enter, to separate artefacts of either; (3)~the fixture is
identified from the measurement alone; (4)~with the fixture fixed, the
[E]~parameters are calibrated in a loop of prediction, comparison, diagnosis
and update, ending in a Nelder--Mead minimisation \citep{NelderMead1965} of
the cost
\begin{equation}
J = \sum_{\mathrm{1C,\,2C}}
\left[\left(\frac{\Delta T_{\mathrm{rms}}}{\SI{0.5}{\kelvin}}\right)^{2}
+\left(\frac{\Delta T_{\mathrm{peak}}}{\SI{0.5}{\kelvin}}\right)^{2}
+\left(\frac{\Delta Q}{\SI{0.1}{\ampere\hour}}\right)^{2}
+\left(\frac{\Delta V_{\mathrm{rms}}}{\SI{20}{\milli\volt}}\right)^{2}\right],
\label{eq:cost}
\end{equation}
where $\Delta T_{\mathrm{rms}}$ and $\Delta T_{\mathrm{peak}}$ are the
root-mean-square and the peak differences of the can temperature, $\Delta Q$
the difference in the charge passed at the cut-off voltage, $\Delta V_{\mathrm{rms}}$
the root-mean-square difference of the voltage where a voltage trace is
measured, and the scales are the resolution of the measurements; (5)~the
calibrated set is compared with the measurement and the residual it leaves is
examined. Each evaluation of $J$ runs the model at both rates. For the NMC cell
the measurements are the open data of \citet{Ank2023,Ank2023data}; for the LFP
cell, for which no comparable data are published, the author's bench test
(Section~\ref{sec:bench}). The model is that of Section~\ref{sec:fdm} with the
thermal balance of Section~\ref{sec:twonode} in both cases.

\subsection{The NMC cell against the open data}
\label{sec:nmcval}

\textbf{Step 1: transfer and prediction.} The set of Tables~\ref{tab:echem}
and~\ref{tab:thermal} was built from the teardown of \citet{Ank2023} and the
electrochemistry of \citet{Chen2020}, with the windows calibrated on the C/20
capacity (Section~\ref{sec:prov}). The open data used for the comparison are
the can temperature at the cap, centre and bottom during 1C and 2C charges
from the discharged state to 4.2\,V followed by \SI{1800}{\second} of rest,
and the C/3 capacity. The pulse resistances and the impedance of the same
source are not used: they test the kinetic parameters, which the thermal data
do not constrain, and their use is left to the identification follow-up. No voltage
trace of these charges is published; the duration of each constant-current
charge, read from the published current profile, gives the charge passed when
4.2\,V was reached, \NmcQMeasOne\ and \NmcQMeasTwo\,Ah at 1C and 2C
(\NmcSOCMeasOne\ and \NmcSOCMeasTwo\,\% of the C/3 capacity). The transferred
set's prediction for these tests is the dotted curve of
Figure~\ref{fig:nmcank}: it reaches 4.2\,V at \NmcTrOneQ\ and \NmcTrTwoQ\,Ah
and raises the can by \NmcTrOnePk\ and \NmcTrTwoPk\,K against \NmcPkMeasOne\
and \NmcPkMeasTwo\,K measured, mean of the three positions.

\textbf{Step 2: cross-check.} The C/3 capacity, which the window
calibration did not use, is reproduced to \SI{0.02}{\percent}
(Table~\ref{tab:validation}). The cross-check between the published references
concerns the heat: \citet{AboutEnergy2024} report \ValQMeas\,W at 1C and
50\,\% state of charge, computed from a resistance with the reversible heat
excluded and for a cell generation that is not stated, whereas inverting the
temperature measured by \citet{Ank2023} during the 1C charge through the
lumped balance gives 4.2 to \SI{5.7}{\watt}. The two are not consistent, and
the working value $\tau_c = \ValTauWork$\,s of the transferred set, chosen so
that the model's heat and DC resistance stayed near the first, is the choice
the calibration revises.

\textbf{Step 3: the fixture.} The roll node carries \AnkCroll\,J\,K$^{-1}$
and the can node \AnkCcan\,J\,K$^{-1}$, so that the two together equal
$m c_p$ with the measured mass of \SI{355.9}{\gram} \citep{Ank2023} and
$c_p = \SI{884.6}{\joule\per\kilo\gram\per\kelvin}$ transferred with the set.
The can node exchanges heat with the chamber through $hA$, which is not chosen
but obtained from the measured cooling after the two charges: the excess
temperature decays with time constants of 1708 and \SI{1764}{\second}, and
$hA = m c_p/\tau$ with $\tau = \SI{1735}{\second}$ gives
\SI{0.18}{\watt\per\kelvin}, radiation included. This determination fixes the
ratio $C/hA$; the level of the temperature rise then tests the heat. Fitted in
the same way, the model's own cooling gives \AnkTauTail\,s, longer because the
roll cools through the can. The lumped still-air estimate used before these
data were available, $C/hA = \ValTau$\,s with an estimated
$h = \SI{10}{\watt\per\square\metre\per\kelvin}$, now has this measured
counterpart. The second property of the fixture is the heat of the terminal
contacts: during the 2C charge the cap runs \SI{9}{\kelvin} above the bottom
and the gradient collapses within minutes of the cut-off
(Figure~\ref{fig:nmcank}D, shaded), which \citet{Ank2023} attribute to a
contact resistance of about \SI{0.5}{\milli\ohm} per terminal, 2\,W at 44\,A.
This heat does not appear in a voltage sensed at the terminals. It is
represented, as for the LFP cell, by $I^2 R_c$ in the can node, with $R_c$ left
to the calibration.

\textbf{Step 4: calibration.} With the fixture fixed, the transferred set
under-predicts the mean can temperature by \NmcTrOneTrms\ and
\NmcTrTwoTrms\,K root-mean-square and reaches 4.2\,V \NmcTrOnedQabs\ and
\NmcTrTwodQabs\,Ah late, at 1C and 2C. Single-parameter runs showed which
[E]~parameters act on these residuals: the diffusion time $\tau_c$ of the
positive electrode raises the heat and brings the cut-off forward at both
rates; $R_c$ raises the 2C temperature four times as much as the 1C one; the
activation energy of the positive-electrode diffusivity sets how much of the
polarisation survives the \SI{45}{\kelvin} rise of the 2C charge; the anode
window moves the cut-off. The entropic coefficients, with the magnitude of
\citet{ORegan2022}, change the 2C peak by about \SI{1}{\kelvin} and were left
transferred, as were the kinetic parameters, which the thermal data do not
separate from $\tau_c$. The four parameters were then minimised on the cost of
Equation~(\ref{eq:cost}) without its voltage term, $\Delta Q$ taken at 4.2\,V.
$J$ fell from \NmcCostTr\ for the transferred set to \NmcCostCal\ in \NmcNev\
evaluations. The calibrated set has $\tau_c = \NmcTauc$\,s
($D_s = \NmcDs$\,m$^2$\,s$^{-1}$), $E_{a,D_s} = \NmcEa$\,kJ\,mol$^{-1}$,
$R_c = \NmcRc$\,m$\Omega$ and an anode stoichiometry of \NmcThetaA\ at full
charge. Two of these values meet an independent number. $R_c$ corresponds to
\SI{0.5}{\milli\ohm} per terminal, the value \citet{Ank2023} estimated for
their holder and \citet{Allgauer2025} measured in a holder of the same type.
$\tau_c$ is within a factor of 1.2 of the value transferred from
\citet{Chen2020}, \ValTauTrans\,s, which the resistance-based heat of
\citet{AboutEnergy2024} had appeared to exclude; the open thermal data, which
are a direct measurement with a measured boundary, carry more weight than a
figure-only value for an unstated cell generation, and the transferred
diffusivity is restored.

\textbf{Step 5: comparison and residual.} The calibrated set reaches 4.2\,V
at \NmcCalOneQ\ and \NmcCalTwoQ\,Ah against \NmcQMeasOne\ and
\NmcQMeasTwo\,Ah measured, and raises the can by \NmcCalOnePk\ and
\NmcCalTwoPk\,K against \NmcPkMeasOne\ and \NmcPkMeasTwo\,K; it follows the
mean can temperature to within \NmcCalOneTrms\ and \NmcCalTwoTrms\,K
root-mean-square through the charge and the rest (Figure~\ref{fig:nmcank}).
Its mean heat is \NmcQmeanCalOne\,W over the 1C charge and \NmcQmeanCalTwo\,W
over the 2C charge, plus \NmcPcanOne\ and \NmcPcanTwo\,W at the contacts,
against 4.2 to \SI{5.7}{\watt} and 13 to \SI{17}{\watt} inverted from the
measurement. What remains is of two kinds. The 2C residual is largest in the
first ten minutes, where the measured mean rises faster than the model: the
cap, heated by the contacts, leads the mean, and a single can node cannot
carry the \SI{9}{\kelvin} axial gradient. Both rates entered the cost; no test
was held out.

\subsection{The LFP cell on the bench}
\label{sec:bench}

\begin{figure}[h!]
\begin{center}
\includegraphics[width=0.75\textwidth]{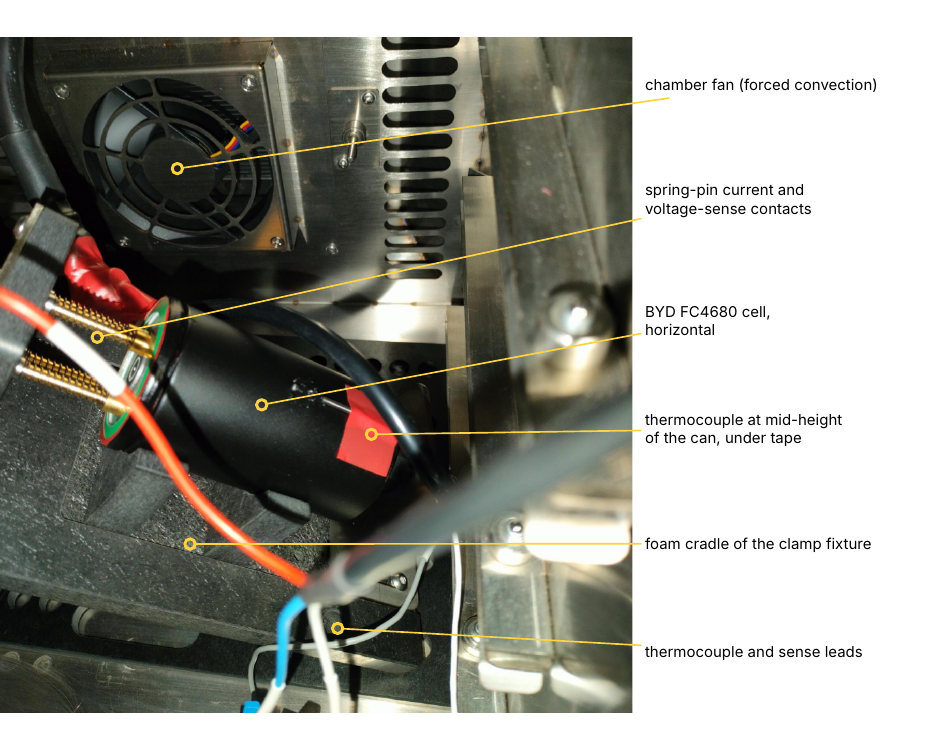}
\end{center}
\caption{The bench setup of the LFP cell: cell c1 lying horizontally in the
foam cradle of the clamp fixture, with the spring-pin current and voltage-sense
contacts at one end, inside the fan-ventilated \SI{25}{\celsius} chamber. For
the discharges of this paper three type~T thermocouples were glued to the can,
\SI{1}{\milli\metre} from each end and at mid-height; the photograph shows
the one at mid-height.}
\label{fig:setup}
\end{figure}

\textbf{Test.} One FC4680 cell (c1) was tested in a horizontal toggle clamp
(Figure~\ref{fig:setup})
with four spring contacts on the positive terminal and a single pointed
contact on the negative terminal, in a chamber held at \SI{25}{\celsius} by
fans. Three type~T thermocouples were glued with epoxy to the can,
\SI{1}{\milli\metre} from each end and at mid-height. After two preconditioning
cycles at 1C, the cell was charged at C/3 to \SI{3.65}{\volt} with a
constant-voltage hold to C/20, rested for \SI{2}{\hour}, discharged at 1C
(\SI{15.29}{\ampere}) to \SI{2.0}{\volt} and rested for \SI{1}{\hour}; the
sequence was repeated with a 2C discharge (\SI{30.59}{\ampere}). The discharges
and the rests after them were recorded at \SI{10}{\hertz}, the rest at
\SI{1}{\hertz}. The predictions of Supplementary Section~S6
were stated before the test. The 10 and \SI{40}{\celsius} runs of the plan
were not carried out.

\begin{figure}[h!]
\begin{center}
\includegraphics[width=0.98\textwidth]{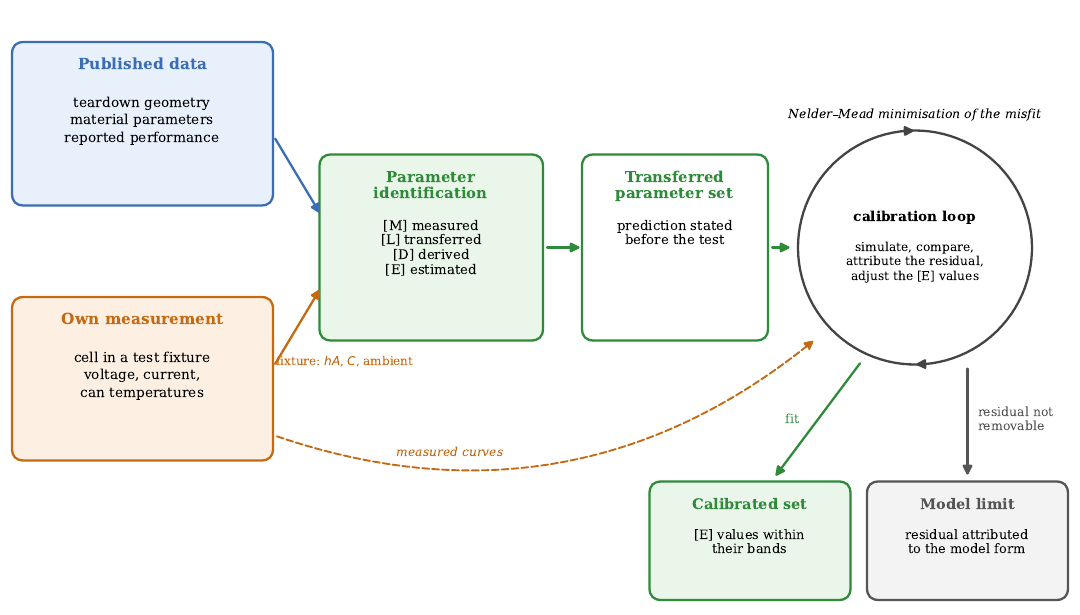}
\end{center}
\caption{Parameterisation and calibration against measurement. Published
data and the measurement feed the identification of
the parameter set, in which every value carries a provenance tag; the fixture
quantities are identified from the measurement alone. The transferred set
gives a prediction that is stated before the test. In the calibration loop
the model is simulated and compared with the measured curves, each residual
is attributed to an artefact of the measurement or to the estimated [E]
values, and the [E] values are adjusted within their bands, the final passes
by Nelder--Mead minimisation of a misfit cost. The loop ends either in a
calibrated set or in a residual that no adjustment removes, which is
attributed to the model form. Section~\ref{sec:validation} applies the
procedure to both cells with the cost of Equation~(\ref{eq:cost}): to the NMC
cell on the open data of \citet{Ank2023}, to the LFP cell on the bench test.}
\label{fig:triangulation}
\end{figure}

\begin{figure}[h!]
\begin{center}
\includegraphics[width=0.72\textwidth]{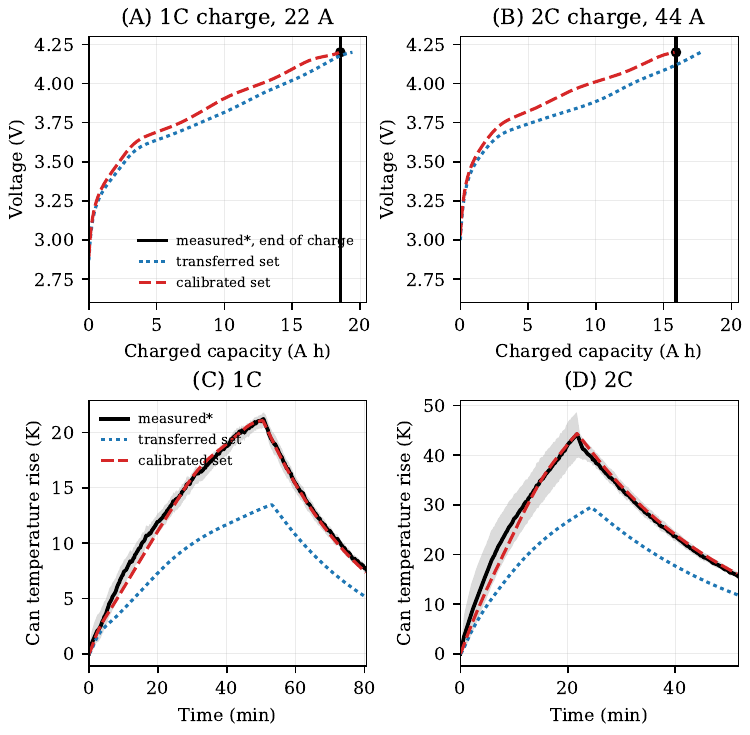}
\end{center}
\caption{\textbf{NMC} cell against the open measurements$^*$ of \citet{Ank2023},
with the transferred set (step~1 of Figure~\ref{fig:triangulation}) and the
calibrated set (step~4). \textbf{(A,B)} Terminal voltage during the
constant-current charge at 1C and 2C from the discharged state, against the
charge passed; no voltage trace is published, and the measured$^*$ point is
the charge at which 4.2\,V was reached, read from the published current
profile. \textbf{(C,D)} Rise of the can temperature through the charge and the
following \SI{1800}{\second} of rest: measured$^*$ mean of the cap, centre and
bottom (their range shaded; it reaches \SI{9}{\kelvin} during the 2C charge),
and the can node of the two-node balance with $hA = \SI{0.18}{\watt\per\kelvin}$
from the measured cooling and the whole-cell heat capacity. Line styles as in (A) in all
panels.
$^*$Measured by \citet{Ank2023}, digitised from their Figure 16.}
\label{fig:nmcank}
\end{figure}

\begin{figure}[h!]
\begin{center}
\includegraphics[width=0.72\textwidth]{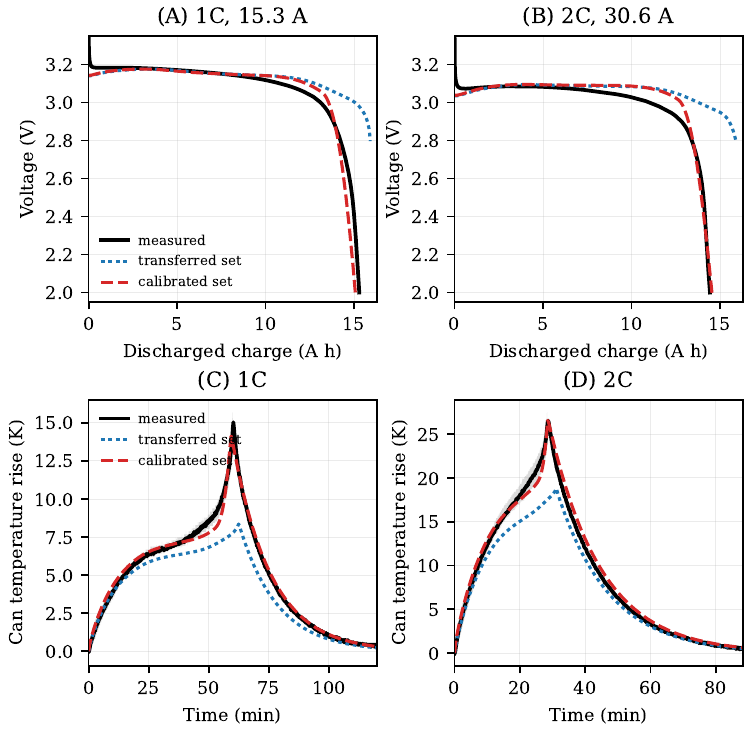}
\end{center}
\caption{\textbf{LFP} cell c1 on the bench, discharges at 1C and 2C from full
charge to \SI{2.0}{\volt} in the \SI{25}{\celsius} chamber (effective ambient
\BenchTamb\,$^\circ$C), against the model with the transferred set (step~1 of
Figure~\ref{fig:triangulation}) and the calibrated set (step~4), both with the
measured
$hA = \BenchhA$\,W\,K$^{-1}$ and heat capacity \BenchC\,J\,K$^{-1}$.
\textbf{(A,B)}~Terminal voltage at 1C and 2C; the transferred set stops at
\SI{2.8}{\volt} with \SI{15.9}{\ampere\hour} delivered.
\textbf{(C,D)}~Rise of the can temperature over the rest temperature, mean of
three thermocouples (their range, typically 0.6\,K at 1C and 1.8\,K at 2C, is
shaded), and the can node of the model, through the discharge
and the following hour of rest. Line styles as in (A) in all panels. The charge delivered to
\SI{2.8}{\volt} and \SI{2.0}{\volt} is compared in Table~\ref{tab:validation}.}
\label{fig:bench}
\end{figure}

\textbf{Procedure.} The LFP set was built and calibrated in the five steps
of Figure~\ref{fig:triangulation}, as the NMC set: transferred from the
teardown of \citet{Liu2025} and the electrochemistry of \citet{AboutEnergyLFP},
each value tagged, with its predictions for the test stated before the test
(Supplementary Section~S6); cross-checked against the
published data; the fixture identified from the bench data alone; the
[E]~parameters calibrated on the cost of Equation~(\ref{eq:cost}), here with
its voltage term, since the voltage is measured; and the residual examined.
Figure~\ref{fig:bench} shows the transferred set of step~1 and the calibrated
set of step~4 against the measurement.

\textbf{Step 2: measurement and cross-check.} The cell delivered \BenchQOne\,Ah at 1C and
\BenchQTwo\,Ah at 2C, and reached \SI{2.8}{\volt} at \BenchQeOne\ and
\BenchQeTwo\,Ah. \citet{Liu2025} report 94.8 and 88.1\,\% of the C/2 capacity to
\SI{2.8}{\volt} at the same rates, 14.2 and \SI{13.2}{\ampere\hour} for their
C/2 capacity of about \SI{15.0}{\ampere\hour}; the measured values lie within
\SI{0.2}{\ampere\hour} of these. The can rose by
\BenchRiseOne\ and \BenchRiseTwo\,K over the rest temperature by
\SI{2.8}{\volt} and peaked 11 to \SI{12}{\second} after cut-off at
\BenchPkOneMeas\ and \BenchPkTwoMeas\,K, mean of the three sensors, with a
maximum of \SI{51.3}{\celsius} at 2C. The sensor near the positive terminal ran
2 to \SI{4}{\kelvin} cooler than the other two at the peak, which is
consistent with the four spring contacts conducting heat into the clamp. The
1C discharge repeats the three preconditioning discharges, made at
\SI{15.0}{\ampere}, to within \SI{0.06}{\ampere\hour} in charge and
\SI{0.5}{\kelvin} in rise at \SI{2.8}{\volt}, of which about \SI{0.4}{\kelvin}
follows from the \SI{2}{\percent} higher current. The cell settles at
\BenchTamb\,$^\circ$C in the \SI{25}{\celsius} chamber after every rest,
including the \SI{2}{\hour} rests, so this is the effective ambient of the
fixture and is used as such. The voltage steps at the start of the 1C
discharge give \SIlist{4.7;7.9;11.7}{\milli\ohm} after 0.1, 1 and
\SI{10}{\second}, nearly the same at 2C; the same steps at the end of the 1C
discharge give \SIlist{7.7;16.5;35}{\milli\ohm}.

The published and the bench data agree on the charge to \SI{2.8}{\volt}
within \SI{0.2}{\ampere\hour} at both rates, so the capacity of the bench cell
is not in doubt. They disagree on the resistance by a factor of 5 to 10: the
area-specific values of \citet{Liu2025}, with the electrode area of
Table~\ref{tab:echem}, give 0.8 to \SI{2.3}{\milli\ohm} against the pulse
resistances above. The normalisation area of the published values is not
stated, and they are not used.

\textbf{Step 3: the fixture.} Over a full cycle the cell
returns to its initial state, so the charge energy less the discharge energy,
\SI{4.78}{\watt\hour} for the 1C cycle, is the heat generated, and all of it
leaves through the fixture. Dividing by the time integral of the measured
excess temperature gives $hA = \BenchhA$\,W\,K$^{-1}$, and the 2C cycle, with
the charge left in the cell accounted for, gives 0.33\,W\,K$^{-1}$. This
determination needs neither the open-circuit voltage nor the heat capacity. It
corresponds to $h = \BenchH$\,W\,m$^{-2}$\,K$^{-1}$ over the whole can, at the
lower edge of the band assumed for the predictions. The cooling after both
discharges decays with a time constant of \BenchTau\,s. With the internal
resistance of Equation~(\ref{eq:rint}) the two-node balance reproduces it for
a heat capacity of \BenchC\,J\,K$^{-1}$; an isothermal roll would require
\SI{315}{\joule\per\kelvin}. Surface sensors cannot separate the two, and the
transferred $m c_p = \LfpCcell$\,J\,K$^{-1}$ lies between them.

\textbf{Step 4: calibration.} At the measured ambient, $hA$ and heat capacity, the
transferred set follows the plateau voltage to within \SI{10}{\milli\volt} at
both rates but places \SI{2.8}{\volt} at \SI{15.9}{\ampere\hour} at both, and
under-predicts the can temperature: \BenchTrmsTrOne\ and \BenchTrmsTrTwo\,K
root-mean-square up to \SI{2.8}{\volt}, at 1C and 2C. It stops at
\SI{2.8}{\volt}, where the particle surfaces saturate and the solver loses its
time step, so it gives no peak.

The calibration is a loop (Figure~\ref{fig:triangulation}): the model is
run, its residuals in voltage, temperature and charge are compared at both
rates, each residual is attributed either to an artefact of a measurement or
to the model, and the treatment of the measurement or the [E] parameters of
the model are updated. The first passes were diagnostic: 14 parameter
directions, each at 1C and most at 2C (Supplementary Table~S8),
identified which parameters act on which residual. They led to four changes.
{\renewcommand{\labelenumi}{(\alph{enumi})}
\begin{enumerate}
\item A heat $I^2R_c$ in the can node that does not enter the terminal
  voltage. With the plateau voltage already matched, the measured heat exceeds
  the model's by about 0.5\,W at 1C and 2\,W at 2C, a ratio of four for a ratio
  of two in current. The heat of the holder contacts outside the voltage sense
  points has this signature; for the NMC cell the same losses were estimated at
  1 to \SI{3}{\watt} at 2C (Section~\ref{sec:validation}).
\item A lower effective diffusivity of the LFP cathode and a lower activation
  energy of it. With the transferred \SI{80}{\kilo\joule\per\mole} no single
  diffusivity places the knee at both rates, because the 2C discharge runs
  about \SI{15}{\kelvin} hotter and diffuses several times faster in the model,
  whereas the cell reaches its knee earlier at 2C.
\item A lower anode stoichiometry at full charge, which corresponds to a loss
  of cyclable lithium relative to the transferred window and sets where the
  graphite is emptied.
\item A floor on the exchange-current density of $2\times10^{-4}\,c_{s,\max}$
  in its concentration factors, which lets the solver follow the voltage below
  the knee to \SI{2.0}{\volt}. It is numerical and has no effect above the knee.
\end{enumerate}}
The parameters of (a) to (c) were then minimised, with the Nelder--Mead
simplex method, on the cost of Equation~(\ref{eq:cost}) without its peak
term, $\Delta V_{\mathrm{rms}}$ and $\Delta T_{\mathrm{rms}}$ taken up to the
measured \SI{2.8}{\volt} point and $\Delta Q$ at that point. Each evaluation
runs the model at both rates. $J$ fell from 11.3 to 3.7 in 50 evaluations.
The minimum, at a factor of 0.37 on the diffusivity, an activation energy of
\SI{25}{\kilo\joule\per\mole}, an anode stoichiometry of 0.731 and
$R_c = \SI{1.3}{\milli\ohm}$, places \SI{2.8}{\volt} at 14.20 and
\SI{13.39}{\ampere\hour} and follows the can temperature to within 0.31 and
\SI{0.53}{\kelvin} up to that point. Below \SI{2.8}{\volt} it does not
(Figure~\ref{fig:bench}, dashed): the cell delivers 1.06 and
\SI{1.03}{\ampere\hour} between \SI{2.8}{\volt} and \SI{2.0}{\volt} at 1C and
2C, this set 0.72 and \SI{1.48}{\ampere\hour}, so the 1C peak is \SI{2.1}{\kelvin}
too low and the 2C peak \SI{1.5}{\kelvin} too high, and after cut-off it
recovers \SI{0.7}{\volt} at once where the cell recovers \SI{0.12}{\volt} in the
first \SI{0.1}{\second} and the rest over 10 to \SI{600}{\second}. Five further
parameter directions around this set, the solid diffusivity of the anode, the
electrolyte diffusivity, the floor on the exchange current, the anode window and
a Nernstian term at the end of the cathode potential, did not remove this
residual (Supplementary Table~S8).

\textbf{The end-of-lithiation potential.} The residual below the knee was
traced to the open-circuit potential of the cathode. The transferred fit of
\citet{AboutEnergyLFP} is valid up to $\theta = 0.950$, and the set
extrapolates it linearly beyond that (Supplementary
Section~S1), which is nearly flat. The fit itself falls
from \SI{3.39}{\volt} at $\theta = 0.95$ to \SI{2.91}{\volt} at 0.99 and
\SI{1.94}{\volt} at 1.0, which is the end of lithiation of LiFePO$_4$. The
guard removes the feature that ends the discharge, so in the model the end was
set by the anode emptying, which does not move with the rate, whereas in the
cell it is set by the cathode and moves with the rate. With the fit used to
$\theta = 1$, and the anode window reopened, the model ends 1C and 2C
\SI{0.84}{\ampere\hour} apart, as measured (\SI{0.86}{\ampere\hour}), but
falls from \SI{2.8}{\volt} to \SI{2.0}{\volt} within \SI{0.2}{\ampere\hour},
where the cell takes \SI{1}{\ampere\hour}. The drop was therefore given a
width: the guarded fit is kept and a term
$-A\,[(\theta - 1 + w)/w]^{n}$ for $\theta > 1 - w$ is added, with
$A = \SI{1.4}{\volt}$ fixed by the fit's value at $\theta = 1$, and the width
$w$ and exponent $n$ left free. This is change (e); it is an effective
description of a drop whose shape, in a real electrode, follows from the
distribution of particle states, and it is tagged [E].

The cost was then extended to the full Equation~(\ref{eq:cost}), with the
peak term and the charge at \SI{2.0}{\volt} added to the charge at
\SI{2.8}{\volt}, and minimised over the six parameters of (a) to (c) and
(e), each evaluation now running to \SI{2.0}{\volt}. On this cost the set above scores
64; the minimum, reached in 45 evaluations, scores 21.4 and is the calibrated
set: a factor of \BenchDsf\ on the diffusivity, $\tau_c = \BenchTauc$\,s, an
activation energy of \BenchEa\,kJ\,mol$^{-1}$, an anode stoichiometry of
\BenchThetaA, $R_c = \BenchRc$\,m$\Omega$, $w = \BenchOcpW$ and
$n = \BenchOcpN$.

\textbf{Step 5: comparison and residual.} The calibrated set places
\SI{2.8}{\volt} at \BenchQmOne\ and \BenchQmTwo\,Ah against \BenchQeOne\ and
\BenchQeTwo\,Ah measured, ends the discharge at \BenchQfOne\ and
\BenchQfTwo\,Ah against \BenchQOne\ and \BenchQTwo\,Ah, and peaks at
\BenchPkOneMod\ and \BenchPkTwoMod\,K against \BenchPkOneMeas\ and
\BenchPkTwoMeas\,K (Figure~\ref{fig:bench}). The can temperature is followed
to within \BenchTrmsOne\ and \BenchTrmsTwo\,K root-mean-square over the whole
discharge, and the voltage to within 27 and \SI{46}{\milli\volt} up to
\SI{2.8}{\volt}. After cut-off the voltage now recovers over the measured range of
time, \SI{2.81}{\volt} against \SI{2.83}{\volt} after \SI{100}{\second} at 1C,
and settles at \SI{3.17}{\volt} against \SI{3.21}{\volt} after 2C
(Table~\ref{tab:validation}). The shape of the recovery differs: the model
recovers 0.33 and \SI{0.43}{\volt} at once at 1C and 2C, then little until
about \SI{10}{\second}, then most of the rest between 10 and
\SI{300}{\second}; the cell recovers 0.15 and \SI{0.21}{\volt} within
\SI{0.1}{\second} and the rest at a nearly constant rate per decade of time
from 0.1 to \SI{1000}{\second}. What remains is a gradual decline of the measured
voltage before the knee that the model does not follow: at 2C the cell is at
\SI{3.0}{\volt} after \SI{11.0}{\ampere\hour}, the model after
\SI{12.8}{\ampere\hour}, and the model is up to \SI{90}{\milli\volt} too high
between 10 and \SI{13}{\ampere\hour}; at 1C the discharge ends
\SI{0.2}{\ampere\hour} early and the peak is \SI{0.9}{\kelvin} low.
Section~\ref{sec:discussion} discusses this residual. The details, the
acceptance criteria stated before the test and the calibration runs are in
Supplementary Section~S6.

\textbf{Estimation methods.} The time constants of the cooling are linear
least-squares fits of the logarithm of the excess temperature,
300 to \SI{3300}{\second} after cut-off, for each sensor and for their mean.
$hA$ is the ratio of the heat over a full cycle to the time integral of the
excess temperature, evaluated by the trapezoidal rule, and the two cycles give
values within \SI{5}{\percent} of each other. The model parameters are the
minimiser of the extended cost of Equation~(\ref{eq:cost}) found by the
Nelder--Mead simplex method \citep{NelderMead1965}, as implemented in
\textsc{SciPy} \citep{Virtanen2020}; it needs no derivatives of the model and
is a local search, started from the diagnostic set. No confidence intervals are given for them:
the repeatability of four 1C discharges of the same cell,
\SI{0.06}{\ampere\hour} in charge and \SI{0.5}{\kelvin} in rise at
\SI{2.8}{\volt}, sets the scale below which differences are not resolved, and
a statistical estimation with uncertainties, of the kind outlined for the NMC
cell in Section~\ref{sec:validation}, requires more than one cell and is left
to the identification follow-up.

The comparison tests the cell-level behaviour of the parameter sets. The
distribution of temperature inside the roll is not resolved by the two-node
balance and is not tested here.

\section{Results}
\label{sec:results}

Tables~\ref{tab:echem} and~\ref{tab:thermal} give the parameters of both cells,
grouped into geometric and electrochemical parameters and thermal parameters.
The complete sets, with the source and tolerance of every value, are in the
Supplementary Material.

\textbf{Electrochemical parameters.} The two cells share the graphite anode
and the electrolyte, and their anode particle radii, rate constants and
electrolyte properties are close. They differ in the cathode. The LiFePO$_4$
particles are ten times smaller than the NMC811 particles, and the diffusion
times of the two cathodes, both calibrated, are of the same order:
\BenchTauc\,s for the LFP cathode, an effective quantity for a two-phase
material, and \NmcTauc\,s for the NMC811 cathode, close to the value
transferred from \citet{Chen2020}. The LFP
coatings are thinner and its winding is longer, which gives the LFP cell
\SI{62}{\percent} more electrode area and \SI{70}{\percent} of the capacity of
the NMC cell.

\textbf{Thermal parameters.} The layer properties of the two cells are similar.
The through-plane conductivity of the roll, which sets the difference between
core and surface temperature, is 1.30 and \SI{1.05}{\watt\per\metre\per\kelvin},
and the heat capacities of the rolls differ by \SI{8}{\percent}. The thermal
behaviour of the format is dominated by quantities that do not depend on the
chemistry: the can, the end hardware and the heat-transfer coefficient at the
can surface. For the LFP cell this coefficient and the heat capacity are
measured in the bench fixture; for the NMC cell the coefficient is taken from
the cooling tails of \citet{Ank2023}. In both fixtures a contact heat outside
the voltage path, \NmcRc\ and \BenchRc\,m$\Omega$, is identified from the
temperature at two rates. The chemistry changes the
heat that is generated, and the current collection changes where it is
generated.

\textbf{Release.} Both parameter sets are released in the Battery Parameter
eXchange format \citep{BPX2022}, as version~1.2, which carries the calibrated
values of both cells (Section~\ref{sec:validation}) and the corrections of version~1.1, verified by loading each file into
\textsc{PyBaMM} \citep{Sulzer2021}, and as a table with one row per parameter
and the provenance tag of each value, organised in the three metadata tiers of
\citet{ward2022} (Supplementary Section~S4).

\section{Discussion}
\label{sec:discussion}

\subsection{The end of a fast discharge}
\label{sec:eod}

The end of discharge is where the LFP cell reaches its highest temperature,
and it is the part of the test that the transferred set described worst. Two
findings follow from the calibration.

First, most of the residual was an artefact of the transfer, not of the model
form. The transferred open-circuit potential of the cathode is a fit whose
validity window ends at $\theta = 0.950$, and the parameter set extrapolated it
linearly beyond that, as it does for every fitted potential. For LiFePO$_4$
the end of lithiation lies just beyond that window, and the extrapolation
removed the \SI{1.4}{\volt} drop that ends the discharge. With the drop in
place, the end moves with the rate as measured, the peaks are reproduced
within \SI{0.9}{\kelvin}, and the relaxation after cut-off, which the set
without the drop recovered at once, follows the measured time scale
(Section~\ref{sec:bench}). Parameter changes in five other directions had
not found this, because the error was in a function, not in a value. The same
guard protects the anode potential at $\theta = 0.823$, and the NMC potentials
at the edges of their windows; a potential that is transferred as a fit should
be checked at the ends of the stoichiometry range the model will reach.

Second, the drop had to be given a width. The fit falls from \SI{2.8}{\volt}
to \SI{2.0}{\volt} within $\Delta\theta = 0.01$, which is \SI{0.2}{\ampere\hour}
of this cell, and the cell takes \SI{1}{\ampere\hour}; the calibrated width is
$w = \BenchOcpW$ with exponent \BenchOcpN. In a model with one representative
particle per control volume, the whole electrode reaches the end of lithiation
at once. In a porous LiFePO$_4$ electrode, particles of different size fill in
sequence at nearly the same potential, and the reaction concentrates in a front
that moves through the electrode \citep{Castle2022}; the end of discharge is
then spread over the charge that the last particles still accept. Models that
describe the cathode as an ensemble of phase-separating particles reproduce
the hysteresis and the path dependence that follow from this
\citep{Foster2025,Clarke2026}. The width identified here is an effective
description of that spread, and the residual that remains has the same
signature: the measured voltage declines gradually from about
\SI{10}{\ampere\hour}, by up to \SI{90}{\milli\volt} more than the model at 2C,
before the knee. The calibrated activation energy of the effective
diffusivity, \BenchEa\,kJ\,mol$^{-1}$ against the transferred 80, is a further
sign that the single diffusivity absorbs a mechanism that is not diffusion.
Whether an ensemble description removes the remaining residual with the
parameters found here above the knee can be tested on the same data. The
quartic particle profile already used here changes the response to steps in
current, not the particle ensemble.

\subsection{Heat outside the jelly roll}

For both cells a part of the heat is generated at the terminal contacts of the
holder: \NmcRc\,m$\Omega$ for the NMC cell in the holder of \citet{Ank2023},
\BenchRc\,m$\Omega$ for the LFP cell in the bench clamp. It does not appear
in a voltage sensed at the cell terminals but raises the can temperature by
several kelvin at 2C. Both calibrations identified it from the ratio of the
heat at two currents, which requires $hA$ to be known independently: from the
cooling tails and the heat capacity for the NMC cell, from the energy balance
over a cycle for the LFP cell. For the NMC cell the value agrees with the
\SI{0.5}{\milli\ohm} per terminal that \citet{Ank2023} estimated and
\citet{Allgauer2025} measured in a holder of the same type. The contact heat
enters at one end of the can, and the axial gradient it produces, \SI{9}{\kelvin}
at 2C on the NMC can, is the largest residual a single can node leaves.
Measuring the voltage drop across each contact would give $R_c$ directly.

\subsection{Conduction and relaxation}

The model uses one diffusion time per electrode, identified from the knee
under current. With the end-of-lithiation drop in place the same diffusion
time also describes the relaxation after the discharges in its range and its
end point, to within \SI{0.04}{\volt} after an hour at both rates
(Table~\ref{tab:validation}); without it the model relaxed at once and to the
wrong voltage. The relaxation data were not used in the calibration, so this
is a check of the diffusion time outside the data that set it. The shape of
the recovery is where the two still differ, and the difference has a
mechanism. At cut-off the model's overpotential has three parts: an ohmic
part, a kinetic part, and a concentration part. The first two vanish at once
when the current stops, and the kinetic part is large at the end of discharge
because the exchange-current density falls with $(c_{s,\max}-c_{se})^{1/2}$ as
the surface saturates; together they are the 0.33 and \SI{0.43}{\volt} that
the model recovers immediately. The concentration part then relaxes with the
single time constant of the quartic profile, $r_s^2/(30 D_s)$, about
\SI{160}{\second} at the calibrated diffusivity, which is the step between 10
and \SI{300}{\second}. The cell recovers only its ohmic drop at once, 0.15
and \SI{0.21}{\volt}, close to $I$ times the \SI{1}{\second} resistance, and
the rest at a nearly constant rate per decade of time over four decades, which
is the signature of a distribution of relaxation times rather than one.

Both differences follow from the representative particle, and the quartic
profile does not remove them. The profile adds one state per control volume,
the volume-averaged concentration gradient, and so changes how the surface of
a particle responds to a step in current; it still gives every particle in a
control volume the same radius, the same surface concentration and the same
relaxation time. At cut-off, that one surface is saturated and the kinetic
overpotential of the whole electrode collapses when the current stops. In an
electrode with a distribution of particle sizes, only the smallest particles
are saturated at cut-off; the larger ones still accept lithium at a lower
overpotential, so the instantaneous recovery of the cell is smaller and the
rest is spread over the relaxation times of the distribution. The same
distribution widens the end-of-lithiation drop (Section~\ref{sec:eod}). A
TDFN whose cathode is an ensemble of phase-separating particles
\citep{Foster2025,Clarke2026}, which we refer to as EP-TDFN, carries that
distribution and is expected to reproduce the instantaneous recovery, the
spread of the relaxation and the width of the drop with the parameters found
here above the knee; this is a hypothesis to be tested on the same data, not
a result of this work. The parameters that govern conduction and those that govern relaxation
are not in general the same, and neither needs to be constant over the state
of charge; here they are consistent at the resolution of one cell. The same applies to the NMC cell, whose cathode consists of
polycrystalline secondary particles of 3 to \SI{16}{\micro\metre}
\citep{Ank2023}, so a single particle radius is also an approximation there.

\subsection{Transferability of the parameter sets}

The parameter sets combine the electrochemistry with the geometry, the
inventory and the current collection of each cell, and several values are
effective: the cathode diffusion times and their activation energies, the
stoichiometry windows, the contact heats and the heat-transfer coefficients of
the fixtures.
These absorb effects of the cell design, the test fixture and the
one-dimensional description of the electrode. The sets are therefore
parameter sets of these two cells in this model, not of the materials, and
they should not be transferred to another format, another fixture or another
model without re-identification of the values tagged [E]. The tags show which
values these are. The identified values do not depend on the grid at the
resolution used (Section~\ref{sec:discretisation}).

\section{Conclusion}

Complete parameter sets have been identified for two 4680 cells, a tabless
NMC811/graphite cell and the multi-tab BYD LFP/graphite cell, with the
provenance of every value recorded. The sets are checked against the mass, the
capacity and the winding of the assembled cells. The mass residuals,
\NmcMassErr\,\% and \BydMassErr\,\%, have opposite signs and locate the NMC
shortfall in hardware that the layer model does not contain. For the LFP cell,
three quantities that no source reports, the cathode porosity, the can wall and
the separator thickness, are determined from the mass, capacity and volume
relations together.

The model couples a DFN of the electrode stack to a two-node thermal balance
whose heat capacities and internal resistance follow from the parameter set,
and whose boundary is taken from the measured cooling of the test fixture. The
current collection enters through the series resistance: between
the tabless and the multi-tab cell the collection resistance changes by a
factor of forty, from negligible to comparable with the can wall.

The heat source of \citet{GuWang2000} is completed by evaluating the reaction
heat at the open-circuit potential of the particle-mean concentration, which
restores the heat of mixing; with the printed form the heat is nearly
independent of the solid diffusivity.

Both sets are calibrated by the same procedure: the fixture is identified
from the measurement alone, and the effective parameters, the cathode
diffusion time, its activation energy, the anode window and a contact heat
outside the voltage path, are adjusted by Nelder--Mead minimisation of a
normalised least-squares cost on the voltage and the can temperature at 1C and
2C. For the NMC cell, against the open charge test of \citet{Ank2023} with
the fixture coefficient taken from the measured cooling, the transferred set
under-predicts the can temperature by 8 and \SI{15}{\kelvin} at the end of
the 1C and 2C charges and reaches 4.2\,V late. The calibrated set follows the
mean can temperature to within \NmcCalOneTrms\ and \NmcCalTwoTrms\,K and
the charge at 4.2\,V to within \SI{0.1}{\ampere\hour}. Its contact heat,
\NmcRc\,m$\Omega$, equals the value the authors estimated for their holder,
and its diffusion time, \NmcTauc\,s, is within a factor of 1.2 of the value
transferred from \citet{Chen2020}, which a resistance-based heat value had
appeared to exclude; a direct thermal measurement with a measured boundary
carries more weight than that value. The kinetic parameters, which the
thermal data do not constrain, are left to be identified from the pulse and
impedance data of the same source in a follow-up.

For the LFP cell, the bench test at 1C and 2C gives the fixture conductance
from the energy balance over a cycle, \BenchhA\,W\,K$^{-1}$, and the heat
capacity from the cooling, \BenchC\,J\,K$^{-1}$. The calibration identifies
a contact heat of \BenchRc\,m$\Omega$ outside the voltage path, an effective
cathode diffusivity with a low activation energy, and the anode window. With
these the model follows the can temperature to within 0.3 and \SI{0.5}{\kelvin}
up to the knee but misses the end of discharge, and the cause is the
transferred open-circuit potential of the cathode: its fit ends at a
stoichiometry of 0.95 and the set extrapolated it flat, removing the
end-of-lithiation drop. With the drop restored and its width identified, the
calibrated set follows the can temperature to within \BenchTrmsOne\ and
\BenchTrmsTwo\,K over the whole discharge, reproduces the peaks to within
\SI{0.9}{\kelvin} and the relaxation after cut-off. The residual that remains,
a gradual decline of the voltage before the knee at 2C, is of the kind that a
distribution of particle states produces and a single representative particle
does not; whether an ensemble description removes it can be tested on the
same data.

\section*{Appendix A. The complete model}
\label{app:model}

Table~\ref{tab:equations} lists all equations of the TDFN. Those not
discussed in Section~\ref{sec:fdm} are the textbook balances of the DFN model
\citep{Doyle1993}: conservation of lithium and of charge in the electrolyte
(E1) and in the solid (E2), coupled by the interfacial reaction current
$j^{\mathrm{Li}} = a_s i_n$ with Butler--Volmer kinetics (E4). The boundary
conditions are no flux of lithium through the foils, no electrolyte current
into the foils, the applied current through the aluminium foil, and
continuity of all fluxes at the electrode--separator interfaces.

\subsection*{A.1 Coupling of the equations}
\label{sec:coupling}

Figure~\ref{fig:couple} can be read in eight steps:
\begin{enumerate}\itemsep0pt
\item The applied current enters the solid (E2) at the aluminium foil; it also
  sets the series drop at the terminal (E9).
\item The solid and the electrolyte (E1, E2) pass the potentials
  $\varphi_s$, $\varphi_e$ and the electrolyte concentration $c_e$ to the
  interface (E4).
\item The interface turns the overpotential into the reaction current
  $j^{\mathrm{Li}}$, which it returns to E1 and E2 as the source of their
  charge and lithium balances.
\item The reaction flux $N$ and the mean concentration $\bar c_s$ from E2 drive
  the particle (E3), which returns the surface stoichiometry $\theta$.
\item The stoichiometry sets the open-circuit potentials (E5, E6), which return
  $U$ to the overpotential of E4. This closes the electrochemical loop.
\item The terminal voltage (E9) is the difference of the solid potentials at
  the two foils minus the series drop.
\item The heat (E7) is collected from the reaction current and overpotential
  of E4, the ohmic losses of E1 and E2, and the open-circuit potential at the
  mean stoichiometry with its temperature coefficient.
\item The two-node balance (E8) adds the $I^2R_{\mathrm{series}}$ heat, turns
  the heat into the roll and can temperatures, and the roll temperature sets
  the transport and kinetic coefficients of E1 to E4.
\end{enumerate}
The steps give the direction of each dependence, not an order of
computation: all equations are solved together within each time step
(Section~\ref{sec:discretisation}), so the loops of steps 3, 5 and 8 close
within the step.

\section*{Conflict of Interest Statement}

WP is employed by NXP Semiconductors Austria GmbH \& Co KG. The remaining
authors declare that the research was conducted in the absence of any
commercial or financial relationships that could be construed as a potential
conflict of interest.

\section*{Author Contributions}

RT: conceptualisation, methodology, software, validation, formal analysis,
data curation, writing (original draft), visualisation. NAW, BB and YZ: bench
test of the LFP cell (investigation, resources, data curation). WP: review of
the model and the manuscript (writing, review and editing). All authors read
and approved the submitted version.

\section*{Funding}

This work was supported by the Villum Investigator Smart Battery Project
(grant 22860) at Aalborg University.

\section*{Generative AI Statement}

Generative AI (Claude, Anthropic) was used to assist with writing and editing
the text, writing and checking code, and preparing figures. The authors
reviewed and edited all generated content and take full responsibility for
the manuscript.

\section*{Data Availability Statement}

The parameter sets for both cells and the closure checks are openly
available. Parameters are released as Battery Parameter eXchange JSON
\citep{BPX2022} for direct loading into \textsc{PyBaMM} and other BPX-aware
tools, and as tidy CSV and JSON with provenance tags. The CSV is organised into the primary,
secondary and tertiary metadata tiers of \citet{ward2022}, with the closure
checks released as quality control on the tertiary tier. The LFP cell's BPX
file is released under CC-BY-SA~4.0 because it incorporates electrochemistry
from \citet{AboutEnergyLFP}; every other artefact is CC~BY~4.0. The geometric and
inventory data for the LFP cell derive from the published teardown of
\citet{Liu2025}; no manufacturer-confidential material is reproduced or relied
upon. The measurements used for the NMC comparison are those of
\citet{Ank2023}, openly available from mediaTUM \citep{Ank2023data}
(\url{https://doi.org/10.14459/2023mp1725661}, CC~BY~4.0); the values used here
were digitised from the published figures and are included in the release. The
bench data of the LFP cell (cell c1, 1C and 2C discharges at \SI{25}{\celsius},
voltage, current and three surface temperatures) and the calibration logs are also included.

The data of this paper are available at
\url{https://github.com/RemusTeodorescu/4680-cell-models}; a versioned
archive with a DOI will be deposited on Zenodo on acceptance.

\section*{Supplementary Material}

The Supplementary Material gives the numerical implementation of the solver,
the constitutive relations, the source and tolerance of every parameter, the
machine-readable release, the validation calculations, the bench test of the
LFP cell with its predictions, results and calibration runs, and the complete parameter sets of both cells.
Every supplementary table is generated from the parameter objects of the
solvers.

\bibliographystyle{Frontiers-Harvard}
\bibliography{references}

\section*{Figures}

\clearpage
\section*{Tables}

\begin{table}[h!]
\caption{Geometric and electrochemical parameters of the two cells. Tags as in Section~\ref{sec:prov}, NMC / LFP. The complete sets, with sources and tolerances, are in Supplementary Sections~S3, S7 and S8. Open-circuit potentials: \citet{Chen2020} (NMC) and \citet{AboutEnergyLFP} (LFP); electrolyte transport: \citet{Nyman2008}. Values marked $^{\dagger}$ are calibrated, the NMC set on the open charge test of \citet{Ank2023} (Section~\ref{sec:nmcval}) and the LFP set on the bench test (Section~\ref{sec:bench}); the values of the transferred sets are 0.701, \SI{700}{\second} and \SI{25}{\kilo\joule\per\mole} (NMC) and 0.823, \SI{3637}{\second} and \SI{80}{\kilo\joule\per\mole} (LFP); the end-of-lithiation drop is the term added to the transferred LFP potential in Section~\ref{sec:bench}.}\label{tab:echem}
\small
\begin{tabular*}{\textwidth}{@{\extracolsep{\fill}}p{78mm}lrrl@{}}
\toprule
Quantity & Unit & NMC & LFP & Tags \\
\midrule
\multicolumn{5}{@{}l}{\emph{Cell}}\\
Winding length & \si{\metre} & 3.27 & 5.37 & [M] / [M] \\
Electrode height & \si{\milli\metre} & 70.0 & 69.0 & [M] / [M] \\
Roll core radius & \si{\milli\metre} & 2.50 & 3.16 & [M] / [M] \\
Roll outer radius & \si{\milli\metre} & 22.2 & 22.7 & [M] / [D] \\
Cu foil thickness & \si{\micro\metre} & 8.00 & 10.0 & [M] / [M] \\
Al foil thickness & \si{\micro\metre} & 10.0 & 10.0 & [M] / [M] \\
Capacity & \si{\ampere\hour} & 22.4 & 15.5 & [D] / [D] \\
N/P ratio, full stoichiometry ranges & --- & 0.929 & 1.18 & [D] / [D] \\
Hardware series resistance & \si{\milli\ohm} & 0.500 & 0.500 & [E] / [E] \\
Holder contact resistance, heat only & \si{\milli\ohm} & \NmcRc$^{\dagger}$ & 1.34$^{\dagger}$ & [E] / [E] \\
\multicolumn{5}{@{}l}{\emph{Negative electrode (graphite)}}\\
Coating thickness, $L_-$ & \si{\micro\metre} & 126 & 54.5 & [M] / [M] \\
Porosity, $\varepsilon_e$ & --- & 0.297 & 0.319 & [D] / [D] \\
Particle radius, $r_s$ & \si{\micro\metre} & 5.86 & 4.80 & [L] / [L] \\
Maximum concentration, $c_{s,\max}$ & \si{\mole\per\cubic\metre} & 30\,555 & 31\,400 & [L] / [L] \\
Stoichiometry at 100\,\% SOC, $\theta_0$ & --- & \NmcThetaA$^{\dagger}$ & 0.796$^{\dagger}$ & [E] / [E] \\
Stoichiometry at 0\,\% SOC, $\theta_{\min}$ & --- & 0.0274 & 0.00163 & [D] / [L] \\
Solid diffusivity, $D_s$ & \si{\square\metre\per\second} & $3.30\times10^{-14}$ & $9.60\times10^{-15}$ & [L] / [L] \\
Rate constant, $m_{\mathrm{ref}}$ & \si{\ampere\metre\tothe{2.5}\per\mole\tothe{1.5}} & $6.48\times10^{-7}$ & $6.68\times10^{-7}$ & [L] / [L] \\
Solid conductivity, $\sigma$ & \si{\siemens\per\metre} & 215 & 7.46 & [L] / [L] \\
\multicolumn{5}{@{}l}{\emph{Separator}}\\
Thickness, $L_{\mathrm{sep}}$ & \si{\micro\metre} & 12.0 & 13.7 & [L] / [D] \\
Porosity, $\varepsilon_e$ & --- & 0.470 & 0.470 & [L] / [L] \\
\multicolumn{5}{@{}l}{\emph{Positive electrode (NMC811 / LiFePO$_4$)}}\\
Coating thickness, $L_+$ & \si{\micro\metre} & 85.0 & 70.0 & [M] / [D] \\
Porosity, $\varepsilon_e$ & --- & 0.335 & 0.331 & [L] / [D] \\
Particle radius, $r_s$ & \si{\micro\metre} & 5.22 & 0.500 & [L] / [L] \\
Maximum concentration, $c_{s,\max}$ & \si{\mole\per\cubic\metre} & 51\,555 & 21\,200 & [L] / [L] \\
Stoichiometry at 100\,\% SOC, $\theta_0$ & --- & 0.264 & 0.0875 & [E] / [L] \\
Stoichiometry at 0\,\% SOC, $\theta_{\max}$ & --- & 0.889 & 0.950 & [D] / [L] \\
Diffusion time, $\tau_c = r_s^2/D_s$ & \si{\second} & \NmcTauc$^{\dagger}$ & 4\,740$^{\dagger}$ & [E] / [E] \\
End-of-lithiation drop, width $w$ / exponent $n$ & --- & --- & 0.131 / 2.6$^{\dagger}$ & --- / [E] \\
Rate constant, $m_{\mathrm{ref}}$ & \si{\ampere\metre\tothe{2.5}\per\mole\tothe{1.5}} & $3.42\times10^{-6}$ & $1.40\times10^{-7}$ & [L] / [L] \\
Solid conductivity, $\sigma$ & \si{\siemens\per\metre} & 0.180 & 0.800 & [L] / [L] \\
\multicolumn{5}{@{}l}{\emph{Electrolyte}}\\
Initial concentration, $c_{e,0}$ & \si{\mole\per\cubic\metre} & 1\,000 & 1\,000 & [L] / [L] \\
Transference number, $t_+^0$ & --- & 0.259 & 0.259 & [L] / [L] \\
\bottomrule
\end{tabular*}
\end{table}

\begin{table}[h!]
\caption{Equations of the TDFN, labelled as in Figure~\ref{fig:couple}. $x$ runs through the repeating unit, $r$ through a particle; $\theta=c_{se}/c_{s,\max}$ and $\bar\theta=\bar c_s/c_{s,\max}$ are the surface and mean stoichiometries. Boundary conditions: no flux of lithium through the foils, no electrolyte current into the foils, the applied current through the aluminium foil, and continuity of all fluxes at the electrode--separator interfaces. The last column gives the equation of the text.}\label{tab:equations}
\small
\renewcommand{\arraystretch}{1.0}\newcommand{\st}{\rule[-2.4ex]{0pt}{6.0ex}}
\setlength{\tabcolsep}{3pt}
\begin{tabular}{@{}p{7mm}>{\raggedright\arraybackslash}p{22mm}>{\raggedright\arraybackslash}p{121mm}l@{}}
\toprule
No. & Element & Equations & Text\\
\midrule
(E1) & Electrolyte & \st$\displaystyle \frac{\partial(\varepsilon_e c_e)}{\partial t}=\frac{\partial}{\partial x}\Big(D_e^\mathrm{eff}\frac{\partial c_e}{\partial x}\Big)+\frac{1-t_+^0}{F}j^\mathrm{Li}$ & --\\
 & & $\displaystyle \frac{\partial}{\partial x}\Big(\kappa^\mathrm{eff}\frac{\partial \varphi_e}{\partial x}+\kappa_D^\mathrm{eff}\frac{\partial \ln c_e}{\partial x}\Big)+j^\mathrm{Li}=0$ & \st--\\
(E2) & Solid & \st$\displaystyle \frac{\partial}{\partial x}\Big(\sigma^\mathrm{eff}\frac{\partial \varphi_s}{\partial x}\Big)-j^\mathrm{Li}=0$,\qquad $\displaystyle \frac{\partial(\varepsilon_s\bar c_s)}{\partial t}=-\frac{j^\mathrm{Li}}{F}$ & --, --\\
(E3) & Particle (quartic) & \st$\displaystyle \frac{\mathrm{d}\bar q}{\mathrm{d}t}=-\frac{30D_s}{r_s^2}\bar q-\frac{45}{2}\frac{N}{r_s^2}$,\qquad $\displaystyle \frac{35D_s}{r_s}(c_{se}-\bar c_s)-8D_s\bar q=-N$ & (\ref{eq:cse})\\
 & & $N=\pm j^\mathrm{Li}/(a_sF)$,\qquad $a_s=3\varepsilon_s/r_s$ & \st\\
(E4) & Interface & \st$\displaystyle j^\mathrm{Li}=a_s i_0\Big[\exp\Big(\frac{\alpha F\eta}{RT}\Big)-\exp\Big(-\frac{\alpha F\eta}{RT}\Big)\Big]$,\qquad $\eta=\varphi_s-\varphi_e-U$ & --\\
 & & $i_0=k\,c_e^{1/2}\big[c_{se}(c_{s,\max}-c_{se})\big]^{1/2}$ & \st\\
(E5) & Anode potential & \st$U=U_-(\theta_-)$, graphite; open-circuit potentials of both cells as in Table~\ref{tab:echem} & Suppl.\\
(E6) & Cathode potential & \st$U=U_+(\theta)$, NMC or LFP; for LFP with the end-of-lithiation drop of Section~\ref{sec:bench} & Suppl.\\
(E7) & Heat & \st$\displaystyle q=j^\mathrm{Li}\Big(\varphi_s-\varphi_e-U(\bar\theta)+T\frac{\partial U}{\partial T}\Big)+\sigma^\mathrm{eff}\Big(\frac{\partial\varphi_s}{\partial x}\Big)^2$ & (\ref{eq:q})\\
 & & $\displaystyle \qquad+\kappa^\mathrm{eff}\Big(\frac{\partial\varphi_e}{\partial x}\Big)^2+\kappa_D^\mathrm{eff}\frac{\partial\ln c_e}{\partial x}\frac{\partial\varphi_e}{\partial x}$ & \st\\
(E8) & Two-node thermal & \st$\displaystyle C_r\frac{\mathrm{d}T_r}{\mathrm{d}t}=\int_\mathrm{roll}q\,\mathrm{d}V+Q_{s,r}-\frac{T_r-T_c}{R_\mathrm{int}}$ & (\ref{eq:energy})\\
 & & $\displaystyle C_c\frac{\mathrm{d}T_c}{\mathrm{d}t}=\frac{T_r-T_c}{R_\mathrm{int}}+Q_{s,c}-hA(T_c-T_\infty)$ & \st\\
 & & $Q_{s,r}+Q_{s,c}=I^2R_\mathrm{series}$, $R_\mathrm{int}$ of Equation~(\ref{eq:rint}) & \st(\ref{eq:rint})\\
(E9) & Terminal & \st$V=\varphi_s(x_\mathrm{Al})-\varphi_s(x_\mathrm{Cu})-IR_\mathrm{series}$;\qquad $D_e$, $\kappa$, $D_s$ and $k$ Arrhenius in $T_r$ & Sec.~\ref{sec:collection}\\
\bottomrule
\end{tabular}
\end{table}

\begin{table}[h!]
\caption{Thermal parameters of the two cells. Layer properties are phase-consistent averages (Equation~\ref{eq:phase}); the effective conductivities of the roll and its heat capacity are derived from them. Tags as in Table~\ref{tab:echem}; $^{\dagger}$ as there. The LFP heat-transfer coefficient and heat capacity are measured in the bench fixture (Section~\ref{sec:bench}), the NMC coefficient in the fixture of \citet{Ank2023} gives $hA = \SI{0.18}{\watt\per\kelvin}$ (Section~\ref{sec:validation}).}\label{tab:thermal}
\small
\begin{tabular*}{\textwidth}{@{\extracolsep{\fill}}p{78mm}lrrl@{}}
\toprule
Quantity & Unit & NMC & LFP & Tags \\
\midrule
\multicolumn{5}{@{}l}{\emph{Layers}}\\
Negative electrode density, $\rho$ & \si{\kilo\gram\per\cubic\metre} & 1\,909 & 1\,887 & [E] / [D] \\
Negative electrode specific heat, $c_p$ & \si{\joule\per\kilo\gram\per\kelvin} & 896 & 908 & [E] / [D] \\
Negative electrode thermal conductivity, $\lambda$ & \si{\watt\per\metre\per\kelvin} & 1.50 & 1.04 & [E] / [E] \\
Separator density, $\rho$ & \si{\kilo\gram\per\cubic\metre} & 1\,056 & 1\,056 & [E] / [D] \\
Separator specific heat, $c_p$ & \si{\joule\per\kilo\gram\per\kelvin} & 1\,693 & 1\,693 & [E] / [D] \\
Separator thermal conductivity, $\lambda$ & \si{\watt\per\metre\per\kelvin} & 0.330 & 0.330 & [E] / [E] \\
Positive electrode density, $\rho$ & \si{\kilo\gram\per\cubic\metre} & 3\,401 & 2\,678 & [E] / [D] \\
Positive electrode specific heat, $c_p$ & \si{\joule\per\kilo\gram\per\kelvin} & 921 & 993 & [E] / [D] \\
Positive electrode thermal conductivity, $\lambda$ & \si{\watt\per\metre\per\kelvin} & 1.50 & 1.50 & [E] / [E] \\
Cu foil density, $\rho$ & \si{\kilo\gram\per\cubic\metre} & 8\,960 & 8\,960 & [L] / [L] \\
Cu foil specific heat, $c_p$ & \si{\joule\per\kilo\gram\per\kelvin} & 385 & 385 & [L] / [L] \\
Cu foil thermal conductivity, $\lambda$ & \si{\watt\per\metre\per\kelvin} & 400 & 401 & [L] / [L] \\
Al foil density, $\rho$ & \si{\kilo\gram\per\cubic\metre} & 2\,700 & 2\,700 & [L] / [L] \\
Al foil specific heat, $c_p$ & \si{\joule\per\kilo\gram\per\kelvin} & 900 & 900 & [L] / [L] \\
Al foil thermal conductivity, $\lambda$ & \si{\watt\per\metre\per\kelvin} & 238 & 237 & [L] / [L] \\
\multicolumn{5}{@{}l}{\emph{Jelly roll (derived)}}\\
Through-plane conductivity & \si{\watt\per\metre\per\kelvin} & 1.31 & 1.05 & [D] / [D] \\
In-plane conductivity & \si{\watt\per\metre\per\kelvin} & 13.4 & 22.6 & [D] / [D] \\
Heat capacity of the roll & \si{\joule\per\kelvin} & 242 & 247 & [D] / [D] \\
Heat capacity of the cell & \si{\joule\per\kelvin} & 315 & 277 & [D] / [M] \\
\multicolumn{5}{@{}l}{\emph{Boundary and temperature dependence}}\\
Can-surface heat-transfer coefficient & \si{\watt\per\square\metre\per\kelvin} & 50.0 & 22.4 & [E] / [M] \\
Entropic coefficient, negative, $\partial U/\partial T$ & \si{\milli\volt\per\kelvin} & 0.0445 & 0.0331 & [L] / [L] \\
Entropic coefficient, positive, $\partial U/\partial T$ & \si{\milli\volt\per\kelvin} & $-$0.0193 & $-$0.0275 & [L] / [L] \\
Activation energy, rate constant $-$, $E_{a,i_0}$ & \si{\kilo\joule\per\mole} & 35.0 & 55.0 & [E] / [L] \\
Activation energy, rate constant $+$, $E_{a,i_0}$ & \si{\kilo\joule\per\mole} & 17.8 & 35.0 & [E] / [L] \\
Activation energy, diffusivity $-$, $E_{a,D_s}$ & \si{\kilo\joule\per\mole} & 30.0 & 30.0 & [E] / [L] \\
Activation energy, diffusivity $+$, $E_{a,D_s}$ & \si{\kilo\joule\per\mole} & \NmcEa$^{\dagger}$ & 18.8$^{\dagger}$ & [E] / [E] \\
Activation energy, electrolyte diffusivity, $E_{a,D_e}$ & \si{\kilo\joule\per\mole} & 17.1 & 17.1 & [E] / [L] \\
Activation energy, electrolyte conductivity, $E_{a,\kappa}$ & \si{\kilo\joule\per\mole} & 17.1 & 17.1 & [E] / [L] \\
\bottomrule
\end{tabular*}
\end{table}

\begin{table}[h!]
\caption{Comparison with measurement. Both cells: the DFN with the two-node thermal balance, quartic particle profile, reaction heat at the open-circuit potential of the particle-mean concentration; \emph{transferred}, the set as predicted before the comparison; \emph{calibrated}, the set after the calibration of Section~\ref{sec:validation}, whose cost contained the can temperature and the charge at cut-off at both rates (and the voltage for the LFP cell). NMC: fixture $hA = 0.18$\,W\,K$^{-1}$ from the measured cooling tails, whole-cell heat capacity, can temperature as the mean of the three measured positions. Reference type: \emph{measured}, a direct measurement; \emph{inverted}, heat obtained from a measured temperature trace through a lumped balance ($\pm$25\,\%); check, the C/3 capacity, not used in the window calibration at C/20. Sources: \citet{Ank2023,Ank2023data} for the charge temperatures and the C/3 capacity; \citet{Liu2025} for the LFP rate retention. LFP: the author's bench test (Section~\ref{sec:bench}), cell c1 at \BenchTamb\,$^\circ$C effective ambient with the measured $hA$ and heat capacity.}
\label{tab:validation}
\begin{center}\footnotesize
\begin{tabular}{@{}l@{\hspace{3mm}}l@{\hspace{3mm}}l@{\hspace{3mm}}l@{\hspace{3mm}}l@{\hspace{3mm}}l@{}}
\toprule
Cell & Quantity & Model & Reference & Difference & Reference type \\
\midrule
NMC & Charge at 4.2\,V, 1C / 2C, transferred & \NmcTrOneQ\ / \NmcTrTwoQ\,\si{\ampere\hour} & \NmcQMeasOne\ / \NmcQMeasTwo\,\si{\ampere\hour} & \NmcTrOnedQ\ / \NmcTrTwodQ\,\si{\ampere\hour} & measured \\
NMC & Charge at 4.2\,V, 1C / 2C, calibrated & \NmcCalOneQ\ / \NmcCalTwoQ\,\si{\ampere\hour} & \NmcQMeasOne\ / \NmcQMeasTwo\,\si{\ampere\hour} & \NmcCalOnedQ\ / \NmcCalTwodQ\,\si{\ampere\hour} & measured \\
NMC & Can temperature, RMS, charge and rest, transferred & --- & --- & \NmcTrOneTrms\ / \NmcTrTwoTrms\,\si{\kelvin} & measured \\
NMC & Can temperature, RMS, charge and rest, calibrated & --- & --- & \NmcCalOneTrms\ / \NmcCalTwoTrms\,\si{\kelvin} & measured \\
NMC & Peak can rise, 1C / 2C, transferred & \NmcTrOnePk\ / \NmcTrTwoPk\,\si{\kelvin} & \NmcPkMeasOne\ / \NmcPkMeasTwo\,\si{\kelvin} & \NmcTrOnedPk\ / \NmcTrTwodPk\,\si{\kelvin} & measured \\
NMC & Peak can rise, 1C / 2C, calibrated & \NmcCalOnePk\ / \NmcCalTwoPk\,\si{\kelvin} & \NmcPkMeasOne\ / \NmcPkMeasTwo\,\si{\kelvin} & \NmcCalOnedPk\ / \NmcCalTwodPk\,\si{\kelvin} & measured \\
NMC & Mean heat, 1C / 2C charge, calibrated, contacts included & \NmcQmeanCalOneTot\ / \NmcQmeanCalTwoTot\,\si{\watt} & 4.2--5.7 / 13--17\,\si{\watt} & --- & inverted \\
NMC & Capacity, C/3 to 2.5\,V & 22.1\,\si{\ampere\hour} & 22.08\,\si{\ampere\hour} & $+$0.02\,\si{\percent} & measured (check) \\
LFP & DC resistance, 1\,s, full charge & 6.18--8.10\,\si{\milli\ohm} & 7.9\,\si{\milli\ohm} & --- & measured; model: inference \\
LFP & Charge to 2.8\,V, 1C / 2C, transferred & 15.9 / 15.9\,\si{\ampere\hour} & 14.24 / 13.41\,\si{\ampere\hour} & $+$12 / $+$19\,\si{\percent} & measured \\
LFP & Can temperature, RMS to 2.8\,V, transferred & --- & --- & \BenchTrmsTrOne\ / \BenchTrmsTrTwo\,\si{\kelvin} & measured \\
LFP & Charge to 2.8\,V, 1C / 2C, calibrated & \BenchQmOne\ / \BenchQmTwo\,\si{\ampere\hour} & 14.24 / 13.41\,\si{\ampere\hour} & $-$1.0 / $+$0.4\,\si{\percent} & measured \\
LFP & Charge to 2.0\,V, 1C / 2C, calibrated & \BenchQfOne\ / \BenchQfTwo\,\si{\ampere\hour} & \BenchQOne\ / \BenchQTwo\,\si{\ampere\hour} & $-$1.4 / $+$0.8\,\si{\percent} & measured \\
LFP & Can temperature, RMS over the discharge, calibrated & --- & --- & \BenchTrmsOne\ / \BenchTrmsTwo\,\si{\kelvin} & measured \\
LFP & Peak can rise, 1C / 2C, calibrated & \BenchPkOneMod\ / \BenchPkTwoMod\,\si{\kelvin} & \BenchPkOneMeas\ / \BenchPkTwoMeas\,\si{\kelvin} & $-$0.9 / 0.0\,\si{\kelvin} & measured \\
LFP & Voltage 1\,h after cut-off, 1C / 2C, calibrated & 3.01 / 3.17\,\si{\volt} & 2.98 / 3.21\,\si{\volt} & $+$0.03 / $-$0.04\,\si{\volt} & measured, not fitted \\
\bottomrule
\end{tabular}
\end{center}
\end{table}

\end{document}